\documentclass[universe,review,accept,pdftex,oneauthor]{Definitions/mdpi} 
\firstpage{1} 
\pubvolume{1}
\issuenum{1}
\articlenumber{0}
\pubyear{2024}
\copyrightyear{2024}
\externaleditor{Academic Editor: Firstname \\Lastname}
\datereceived{ 31 August 2024} 
\daterevised{8 November 2024} 
\dateaccepted{11 November 2024} 
\datepublished{ } 
\hreflink{https://doi.org/} 
\usepackage{amsfonts,amssymb,amsmath,graphicx,multirow,dsfont}
\makeatletter
\let\c@lofdepth\relax
\let\c@lotdepth\relax
\makeatother
\usepackage{subfigure}
\makeatletter
\renewcommand{\@thesubfigure}{\normalsize(\textbf{\alph{subfigure}})}
\makeatother
\graphicspath{{figures/}}
\def\dac{\displaystyle\frac}

\def\[{\left[}
\def\]{\right]}
\def\({\left(}
\def\){\right)}
\def\ot{\leftarrow}
\newcommand{\diag}{\mathop{\rm diag}\nolimits}
\newcommand{\const}{\mathop{\rm const}\nolimits}

\Title{Cosmological Models in Lovelock Gravity: An Overview of Recent Progress}

\TitleCitation{Cosmological Models in Lovelock Gravity: An Overview of Recent Progress}

\Author{{Sergey Pavluchenko} 
 \orcidA{}}

\AuthorNames{Sergey A. Pavluchenko}

\AuthorCitation{{Pavluchenko,} 
 S.A.}

\address[1]{%
{Department of Bioengineering, University of Illinois Urbana-Champaign,} 
 Urbana, IL 61801, USA; sergey.pavluchenko@gmail.com}

\abstract{In the current review, we provide a summary of the recent progress made in the cosmological aspect of extra-dimensional Lovelock gravity. Our review covers a wide variety of particular model/matter source combinations: Einstein--Gauss--Bonnet as well as cubic Lovelock gravities with vacuum, cosmological constant, perfect fluid, spatial curvature, and some of their combinations. Our analysis suggests that it is possible to set constraints on the parameters of the above-mentioned models from the simple requirement of the existence of a smooth transition from the initial singularity to a realistic low-energy regime. Initially, anisotropic space naturally evolves into a configuration with two isotropic subspaces, and if one of these subspaces is three-dimensional and is expanding while another is contracting, we call it realistic compactification. Of course, the process is not devoid of obstacles, and in our paper, we review the results of the compactification occurrence investigation for the above-mentioned models. In particular, for vacuum and $\Lambda$-term EGB models, compactification is not suppressed (but is not the only possible outcome either) if the number of extra dimensions is $D \geqslant 2$; for vacuum cubic Lovelock gravities it is always present (however, cubic Lovelock gravity is defined only for $D \geqslant 3$ number of extra dimensions); for the EGB model with perfect fluid it is present for $D=2$ (we have not considered this model in higher dimensions yet), and in the presence of spatial curvature, the realistic stabilization of extra dimensions is always present (however, such a model is well-defined only in $D \geqslant 4$ number of extra dimensions).}



\begin{document}

\section{Introduction}

By now, mathematical modeling has become an integral part of any quantitative or qualitative study in natural or life sciences and beyond. The very first models in the history of humankind were what we could call ``cosmological models'', as they described the whole world surrounding the ancient scientists---the planets revolving around the Earth. The development of differential calculus by Newton and Leibniz became an important milestone, which boosted both physics and mathematics and, through this, the modeling. However, it took three more centuries and the development of even more powerful tensor calculus for cosmological modeling to become a full-fledged discipline. Nowadays, modeling is an essential part of cosmology and we rely on models for all of its aspects.

The typical workflow in cosmology is the same as in other natural sciences---we formulate a hypothesis, build a model based on this hypothesis, calculate predictions, and confront them with experimental data to decide if the model is viable, and if so, to find a realistic range of parameters. However, there are exceptions from this workflow and one of them is the Very Early Universe. Indeed, there is not much experimental and/or observational evidence from the early stages of the Universe's evolution and most of them are indirect; the earlier the stage, the less evidence we have. In this case, we can rely only on models and use only the most general assumptions of what to expect. 

A good example of such a situation is the topic of the current review---Lovelock gravity in the Early Universe. Lovelock gravity is the generalization of General Relativity (GR) in the following way: we know~\cite{etensor1, etensor2, etensor3} that the Einstein tensor is, in any dimension, the only symmetric and conserved tensor depending only on the metric and its first and second derivatives (with a linear dependence on second derivatives). If we drop the requirement of linear dependence of the second derivatives and use a similar procedure as in GR, we naturally obtain the Lovelock tensor~\cite{Lovelock}. Lovelock gravity manifests itself as a curvature correction to GR: the zeroth (constant) term is just the boundary term associated with the cosmological constant, the first (linear) term is GR, the second (quadratic) term is known as the Gauss--Bonnet (GB) term, and so on. Due to the Lovelock tensor construction procedure, nontrivial---higher-than-GR---contributions from Lovelock gravity manifest themselves only in a higher-than-three number of spatial dimensions, making it extra-dimensional theory. For the Gauss--Bonnet contribution to become nontrivial, we need at least four spatial dimensions, and for the cubic Lovelock term, at least six, and so on. As we just mentioned, Lovelock gravity manifests itself as a curvature correction, so this theory is distinguishable from GR only in the high-curvature regime, which is naturally achieved during the early stages of the Universe's evolution. That is why it is mostly an Early Universe theory. 

A nonlinear theory, Lovelock gravity shares the same curse with other nonlinear theories---unwanted singularities, which we call ``nonstandard singularities'', in contrast with standard cosmological singularities. These nonstandard singularities are referred to as a situation when some of the dynamical variables diverge while others are regular at some finite time; they are considered ``weak'' by Tipler's classification~\cite{Tipler}, and ``type II''
in tje classifications by Kitaura and Wheeler~\cite{KW1, KW2}. Studies demonstrate that they naturally appear during evolution and, being physical singularities (they cause curvature invariants to diverge), cease the Universe's evolution. 

On the other hand, Lovelock gravity is formulated in a greater-than-three number of spatial dimensions, while we clearly live in a three-dimensional {world.}
\endnote{Good proof comes from Newtonian gravity: if defined in more than three spatial dimensions, it has no stable orbits, while the Earth is rotating around the Sun for billions of years.} So, the natural question here is how to connect these two within the same theory. The widely accepted answer is that extra
dimensions are compact---their scale is much smaller than that of three dimensions, so we do not ``sense'' them. This explanation, however, raises another question---how did this compactification occur? In the most elegant way, this compactification should happen naturally during the Universe's evolution. In other words, compactification should be a natural attractor of the system.

So, we have Lovelock gravity, which is formulated in a higher number of spatial dimensions, and we want it to describe our Universe at late times. However, there are next to no observational data from the very early times of the Universe's evolution, so, heeding in mind the previously mentioned potential problems, we want the following natural assumptions to be held: (1) the late-time regime should be low-curvature (GR); (2) the late-time regime should have three expanding dimensions while the remaining should be compact (contracting or constant with the scale much smaller than ``our'' three dimensions); (3) the late-time regime should be achieved smoothly (i.e., without any nonstandard singularities). These three assumptions sound reasonable but we will see that they are sufficient to set constraints on the theory under consideration.

As we already mentioned,
the models under consideration have two features: (a) they are multidimensional and (b) we consider them within Lovelock
gravity. Each of these features has mechanisms that may prevent the model from reaching the realistic regime. Since the
model is multidimensional, it may compactify into wrong spatial splitting---it could be that there is no three-dimensional 
subspace after compactification and the resulting regime is unrealistic---findings on this feature are summarized in
Section~\ref{sec_EGB_aniso}. Another feature is that we consider Lovelock gravity in action, and it brings additional possibilities for
interfering with the evolution---this includes unrealistic late-time asymptotes (such as power-law regimes) or even 
finite-time nonstandard singularities; the rest of the sections are dedicated to the study of these possibilities. 
And since we are living in three-dimensional space, which experiences accelerated expansion, that is what we are
aiming for as a late-time~asymptote.

Let us emphasize that we focus on cosmological behavior in the very early Universe---indeed, Lovelock 
gravity contains terms that are higher-order in curvature and these terms manifest themselves only
when the curvature itself is high, and for cosmological context (especially since we are working mostly with spatially flat 
models) it is usually near the initial singularity
(or recollapse, but this is an undesired outcome). This way, we ensure that if realistic compactification is reached,
the future asymptote is GR and its
behavior is considered the same as in standard GR, as will its manifestations---direct and indirect observables,
qualitative behavior, and so on (BBN predictions, cosmic expansion history, structure formation, etc.). So, to 
put it simply, in the current manuscript, we are summarizing findings on constraints on the parameters of the theory, 
which allows realistic compactification for Lovelock gravity.

The manuscript is organized as follows: first, we shall provide a brief historical outline of the research in cosmological aspects of Lovelock gravity, and after that we will set up a model. With the model being set up, we will consider different Lovelock corrections and matter sources---vacuum and cosmological term in Einstein--Gauss--Bonnet (EGB) gravity (Sections~\ref{sec_EGB_vac} and~\ref{sec_EGB_Lambda}, respectively), vacuum in cubic Lovelock gravity (Section~\ref{sec_L3_vac}), perfect fluid in EGB gravity (Section~\ref{sec_EGB_perf.fluid}), spatial curvature in EGB gravity (Section~\ref{sec_EGB_curv}), and finally, general anisotropic case in EGB gravity (Section~\ref{sec_EGB_aniso}). After that, we discuss the results and draw some~conclusions.

\section{Historical Outline}

\textls[-15]{Surprisingly, the very idea of extra dimensions precedes GR---the first extra-dimensional} theory was formulated by Nordstr\"om in 1914~\cite{Nord1914} and it represents the unification of Nordstr\"om's second gravity theory~\cite{Nord_2grav} with Maxwell's electromagnetism. However, soon after that, Albert Einstein introduced GR~\cite{einst}, which competed with Nordstr\"om's scalar gravity for being the correct theory. This competence was resolved in 1919 in favor of GR: Nordstr\"om's scalar gravity, similarly to most other scalar gravities, predicts no light bending near massive bodies, while observations made during the solar eclipse in 1919 clearly demonstrated that the deflection angle is in agreement with GR predictions.

With that, Nordstr\"om's scalar gravity was abandoned but his brilliant idea survived. Kaluza proposed~\cite{KK1} a similar model but based on GR: five-dimensional Einstein equations could be decomposed into four-dimensional Einstein equations and Maxwell's electromagnetism. His model has one extra dimension and it should be ``curled'' (or compactified) into a circle of a very small size and ``cylindrical conditions'' should be imposed in order to perform dimensional decomposition.
That was followed by Klein who proposed~\cite{KK2, KK3} an interesting quantum mechanical interpretation of this extra dimension and the resulting theory was named Kaluza--Klein, after its founders. It is worth mentioning that the Kaluza--Klein theory united all interactions known at that time. As time passes, new interactions were discovered and it became clear that, to unite all of them, one needs a larger number of extra dimensions. As of now, a few models unite all interactions and M/string theory is one of the most promising of them.

In recent times, theories similar to (or, rather, generalizing) the Kaluza--Klein theory\endnote{Apparently, the original Kaluza--Klein theory is considered incorrect for a number of reasons and follow-up investigations generalize it---see, e.g.,~\cite{Wesson} for review.} are used to address a number of problems in modern physics, with the hierarchy problem being one of the most important of them. In particular, Arkani-Hamed, Dimopoulos, and Dvali~\cite{ADD1, ADD2} proposed a model that involves large extra dimensions with the effective Planck mass at about 1 TeV, while Randall and Sundrum considered a couple of models with warped extra dimensions~\cite{RS1, RS2}, addressing the same hierarchy problem. Observations do not rule out such models but severely constrain them (see, e.g.,~\cite{LED_constr1} for a constraint coming from neutron stars observations by Fermi-LAT telescope), but it is worth mentioning that the models with really large extra dimensions (at least about the length of gravitational waves ($\sim$100 km)) are literally undetectable (see, e.g.,~\cite{LED_constr2} for constraints coming from the analysis of the gravitational waves damping). Extra-dimensional theories, as well as the manifestations of the background theories (M/string), could be detected in particle experiments, but so far, results are discouraging~\cite{ATLAS1, ATLAS2}.

From the gravitational standpoint, one of the distinguishing features of M/string theories is the presence of the curvature-squared corrections in the Lagrangian. Scherk and Schwarz~\cite{sch-sch} noted the presence of the $R^2$ and $R_{\mu \nu} R^{\mu \nu}$ terms in the Lagrangian of the Virasoro--Shapiro model~\cite{VSh1, VSh2}; Candelas et al.~\cite{Candelas_etal} demonstrated the presence of the curvature-squared term of the $R^{\mu \nu \lambda \rho} R_{\mu \nu \lambda \rho}$ type in the low-energy limit of the $E_8 \times E_8$ heterotic superstring theory~\cite{Gross_etal} to compensate the kinetic term of the Yang--Mills field. Later on, Zwiebach showed~\cite{zwiebach} that in order to keep the theory ghost-free, one has to use a specific combination of  quadratic terms known as the Gauss--Bonnet (GB) term, as follows:

$$
L_{GB} = R_{\mu \nu \lambda \rho} R^{\mu \nu \lambda \rho} - 4 R_{\mu \nu} R^{\mu \nu} + R^2.
$$

\noindent {This}  term, originally discovered by Lanczos~\cite{Lanczos1, Lanczos2} (and thus sometimes called the Lanczos term) is an Euler topological invariant in (3+1)-dimensional spacetime, but starting from (4~+~1), and in higher dimensions, it gives a nontrivial contribution to the equations of motion. Zumino~\cite{zumino} extended Zwiebach's analysis with the higher-than-squared curvature terms taken into account; his results suggested that the low-energy limit of the unified theory might have a Lagrangian density represented by a sum of different powers of curvature---then, the earlier-mentioned Lovelock gravity would be a natural candidate for such a theory. 

There are two approaches to the compactification problem, with the first of them being ``spontaneous compactification''~\cite{add_1, Deruelle2} (see also~\cite{add_4} for more cosmology-relevant solutions) and the other is ``dynamical compactification''. Roughly speaking, the difference between these two is as follows: in the former, you make extra dimensions compact ``by hand'' and see if the resulting setup is viable within the gravity theory under consideration, while in the latter, you do not impose small extra dimensions in the beginning, but they become small in the process of the evolution. As ``dynamical compactification'' proposes a more elegant way, it involves different approaches~\cite{add_81, add_82} and setups~\cite{MO1, MO2}. Apart from cosmology,  studies of extra dimensions include the investigation of black holes in Gauss--Bonnet~\cite{BHGB1, BHGB2, BHGB3, BHGB5, BHGB6, BHGB7, BHGB8, BHGB9} and Lovelock~\cite{BHL1, BHL2, BHL3, BHL4, BHL5} gravities, features of the gravitational collapse in these theories~\mbox{\cite{coll1, coll2, coll3}}, general peculiarities of spherical-symmetric solutions~\cite{addn_8}, the formation of shock due to nonlinearity~\cite{shock}, and many others.

The structure of the equations of motion in EGB and more general Lovelock gravity is more complicated than that in GR, as the former are nonlinear theories, unlike the latter. Thus, finding exact solutions within these theories is a very non-trivial task, and to overcome it, one usually applies metric {\it {ansatz}} of some sort. For cosmology, the usual {\it {ans\"atzen}} are power-law and exponential; the former of them resembles the Friedmann stage while the latter, the accelerated expansion nowadays or inflationary stage in the Early \mbox{Universe---the de} Sitter stage.

  The cosmological power-law solution within Lovelock and EGB gravities was studied in~\mbox{\cite{Deruelle1, Deruelle2}} and more recently in~\cite{mpla09, iv1, iv2, grg10, prd09, prd10}, resulting in some understanding of their dynamics. One of the first considerations of the exponential solutions within the considered theories could be found in~\cite{Is86}, while recent works include~\cite{exp_diff1, exp_diff2, exp_diff4, exp_diff5, exp_diff7, exp_diff8, exp_diff9, exp_diff10, exp_diff11, exp_diff12, exp_diff13}; the separate description of the exponential solutions with variable and constant volume was performed in~\cite{CPT1} and~\cite{CST2}, respectively; we could also refer to~\cite{PT} for the discussion about the link between the existence of power-law and exponential solutions and for the discussion about the physical manifestations of different branches of the solutions within these theories. We also offer the full description of the general scheme for finding all possible exponential solutions in arbitrary dimensions and with arbitrary Lovelock contributions taken into account in~\cite{CPT3}. A deeper investigation of the exponential solutions reveals that not all of the solutions found in~\cite{CPT3} are stable~\cite{my15}; a more general approach to the stability of exponential solutions in EGB gravity could be found in~\cite{iv16}, while in~\cite{stab_add1, stab_add2, stab_add3, stab_add4}, some particular cases are described more closely.

The above-mentioned exponential and power-law {\it {ans\"atzen}} are considered as dynamical late-time attractors for the Universe's evolution; however, there could be static attractors as well. As we demonstrated in~\cite{CGP1, CGP2} (and further investigated in~\cite{CGPT}) if we consider extra dimensions to have negative spatial curvature, for some parameter combinations, there exist asymptotic regimes with expanding three and constant-size extra dimensions. Generally, in GR, positive spatial curvature is considered; say, positive spatial curvature could change inflationary asymptotic~\cite{infl1, infl2}. However, in EGB gravity, it is negative curvature which gives rise to the stabilization of extra dimensions; positive curvature could stabilize as well but for a very narrow range of parameters~\cite{our20}; see also~\cite{PT2017, CP21} for some more details on the difference between the cases with positive and negative spatial curvature.

This finalizes our review of preliminary studies dedicated to cosmological solutions in EGB and more general Lovelock gravity. With power-law and exponential solutions being well-described, it is time to find out if they could be realized during the natural evolution of the model, which is exactly what the rest of the paper is dedicated to.

\section{The Model}

As we already mentioned, we work with extra-dimensional theory and we are particularly interested in obtaining compactification in a natural way, meaning that the Universe ends up with three expanding dimensions while extra dimensions remain compact. Then, it is natural to consider the metric {\it {ansatz}} for such a model as spatially flat anisotropic (Bianchi-I-type), as follows:

\begin{equation}
\begin{array}{l}
ds^2 = \diag(-1, a_1^2(t), a_2^2(t), \dots, a_{\tilde{D}}^2(t)),
\end{array} \label{metric1}
\end{equation}

\noindent where $a_i(t)$ is the scale factor corresponding to $i$th spatial dimension and $\tilde{D}$ is the total number of spatial dimensions. To obtain equations of motion, we remind the reader how Lovelock gravity is structured: Lovelock invariants have the form~\cite{Lovelock}

$$
L_n = \frac{1}{2^n}\delta^{i_1 i_2 \dots i_{2n}}_{j_1 j_2 \dots j_{2n}} R^{j_1 j_2}_{i_1 i_2} \dots R^{j_{2n-1} j_{2n}}_{i_{2n-1} i_{2n}},
$$

\noindent where $\delta^{i_1 i_2 \dots i_{2n}}_{j_1 j_2 \dots j_{2n}}$ is the generalized Kronecker delta of the order $2n$. Then, the Lagrangian density has a form

\begin{equation}
\begin{array}{l}
{\cal L}= \sqrt{-g} \sum_n \alpha_n L_n,
\end{array} \label{lagr_gen}
\end{equation}

\noindent with $g$ being the determinant of the metric tensor, $\alpha_n$ being the coupling constant, and the summation over all $n$ in consideration is assumed. The rest of the derivation is quite straightforward (see, e.g.,~\cite{prd09}), and resulting equations, rewritten in terms of Hubble parameters ($H_i = \dot a_i/a_i$), read as follows:

\begin{equation}
\begin{array}{l}
(2n-1) \sum\limits_{k_1 > k_2 > \dots k_{2n}} H_{k_1} H_{k_2}
\dots H_{k_{2n}} = 0
\end{array} \label{constr_gen}
\end{equation}

\noindent for constraint equation and

\begin{equation}
\begin{array}{l}
\sum\limits^D_{\substack{m=1 \\ m\ne i}} \left[ (\dot H_m + H_m^2) \sum\limits_{\substack{\{j_1,\,j_2,\,\dots\,j_{2n-2}\}\ne\{i, m\}\\ j_1 > j_2 > \dots j_{2n-2}}}  H_{j_1} H_{j_2} \dots H_{j_{2n-2}} \right] + \\ \\  \qquad + (2n-1)\sum\limits_{\substack{\{ k_1, k_2, \dots k_{2n}\}\ne i\\ k_1 > k_2 > \dots k_{2n}}} H_{k_1} H_{k_2} \dots H_{k_{2n}} = 0
\end{array} \label{dyn_gen}
\end{equation}

\noindent for dynamical equation corresponding to $i$th spatial coordinate. Please mind that the equations above are obtained for the vacuum case; in the presence of a nontrivial source, the right-hand sides of (\ref{constr_gen}) and (\ref{dyn_gen}) are modified appropriately by adding density and pressure, respectively. 

Our studies demonstrated that $(3 + D)$ spatial splitting\endnote{We use this term to describe the situation when an initially anisotropic space evolves into a 
configuration with two isotropic subspaces; for $(3 + D)$ spatial splitting it would be $(a_1^2, a_2^2, \dots, a_{3+D}^2) \to (a^2, a^2, a^2, b^2, \dots, b^2)$.}
 could be realized under some conditions (see Section~\ref{sec_EGB_aniso} for details), so that we can use this splitting to study the dynamics of the corresponding model. Then, the $(3 + D)$ spatial splitting ($D$ now stands for number of extra dimensions) metric has the form

\begin{equation}
\begin{array}{l}
ds^2 = \diag(-1, a^2(t), a^2(t), a^2(t), b^2(t), \dots, b^2(t)),
\end{array} \label{metric2}
\end{equation}

\noindent and the resulting equations of motion could be obtained from (\ref{constr_gen}) and (\ref{dyn_gen}) by applying metric {\it ansatz} (\ref{metric2}). Then, if we limit ourselves with the most cubic Lovelock contribution ($n=1, 2, 3$ in (\ref{lagr_gen}); that is, the highest Lovelock order we are dealing with in this review) and normalize Gauss--Bonnet and cubic Lovelock couplings to GR coupling ($\alpha_1 \equiv 1$, $\alpha_2 \equiv \alpha$, $\alpha_3 \equiv \beta$), and use notations $H \equiv \dot a/a$, $h \equiv \dot b/b$, dynamical equations could be rewritten as

\begin{equation}
\begin{array}{l}
2 \[ 2 \dot H + 3H^2 + D\dot h + \dac{D(D+1)}{2} h^2 + 2DHh\] + 8\alpha \[ 2\dot H \(DHh + \dac{D(D-1)}{2}h^2 \) + \right. \\
 \\ \left. + D\dot h \(H^2 + 2(D-1)Hh + \dac{(D-1)(D-2)}{2}h^2 \) +
2DH^3h + \dac{D(5D-3)}{2} H^2h^2 + \right. \\
\\ \left. + D^2(D-1) Hh^3 + \dac{(D+1)D(D-1)(D-2)}{8} h^4 \] +  \\ \\
+  144\beta \[ \dot H \(Hh^3 \dac{D(D-1)(D-2)}{3} +
h^4\dac{D(D-1)(D-2)(D-3)}{12} \) + \right. \\ \\
+ \left. D \dot h
\(H^2 h^2 \dac{(D-1)(D-2)}{2} +  Hh^3 \dac{(D-1)(D-2)(D-3)}{3} + \right. \right. \\
\\ + \left. \left.  h^4\dac{(D-1)(D-2)(D-3)(D-4)}{24} \) +  H^3h^3 \dac{D(D-1)(D-2)}{3} + \right. \\ \\
+ \left. H^2h^4 \dac{D(D-1)(D-2)(7D-9)}{24} +  Hh^5 \dac{D^2(D-1)(D-2)(D-3)}{12} + \right. \\ \\
+ \left. h^6 \dac{(D+1)D(D-1)(D-2)(D-3)(D-4)}{144} \]
 - \Lambda=0
\end{array} \label{H_gen}
\end{equation}

\noindent for a dynamical equation that corresponds to $H$
\begin{adjustwidth}{-\extralength}{0cm}
\begin{equation}
\begin{array}{l}
2 \[ 3H^2 + 3DHh + \dac{D(D-1)}{2} h^2 \] + 24\alpha \[ DH^3h + \dac{3D(D-1)}{2}H^2h^2 + \right. \\ \\ \left. + \dac{D(D-1)(D-2)}{2}Hh^3 +  \dac{D(D-1)(D-2)(D-3)}{24}h^4\] +
 720\beta \[ H^3 h^3 \dac{D(D-1)(D-2)}{6} + \right. \\ \\ \left. +
 H^2 h^4 \dac{D(D-1)(D-2)(D-3)}{8} +  Hh^5 \dac{D(D-1)(D-2)(D-3)(D-4)}{40} + \right. \\ \\ \left. + h^6 \dac{D(D-1)(D-2)(D-3)(D-4)(D-5)}{720}    \] = \Lambda
\end{array} \label{con_gen}
\end{equation}
\end{adjustwidth}

\noindent for a constraint equation and

\begin{equation}
\begin{array}{l}
2 \[ 3 \dot H + 6H^2 + (D-1)\dot h + \dac{D(D-1)}{2} h^2 + 3(D-1)Hh\] + 8\alpha \[ 3\dot H \(H^2 + \right. \right. \\
\\ \left. \left. + 2(D-1)Hh +  \dac{(D-1)(D-2)}{2}h^2 \) +  (D-1)\dot h \(3H^2 + 3(D-2)Hh + \right. \right. \\
\\  \left. \left. +
\dac{(D-2)(D-3)}{2}h^2 \) + 3H^4 +  9(D-1)H^3h + 3(D-1)(2D-3) H^2h^2 + \right. \\
\\ \left. + \dac{3(D-1)^2 (D-2)}{2} Hh^3 +   \dac{D(D-1)(D-2)(D-3)}{8} h^4 \] + \\ \\  + 144\beta\[ \dot H \( H^2 h^2 \dac{3(D-1)(D-2)}{2} +  Hh^3(D-1)(D-2)(D-3) + \right. \right. \\
\\ + \left. \left.
h^4 \dac{(D-1)(D-2)(D-3)(D-4)}{8} \)  + (D-1)\dot h \(  H^3 h (D-2) + \right. \right. \\
\\ + \left. \left. H^2h^2 \dac{3(D-2)(D-3)}{2} +  Hh^3 \dac{(D-2)(D-3)(D-4)}{2} + \right. \right. \\
\\  \left. \left. +
h^4 \dac{(D-2)(D-3)(D-4)(D-5)}{24}  \) + H^4 h^2 \dac{3(D-1)(D-2)}{2} + \right. \\ \\ \left. + H^3h^3 \dac{(D-1)(D-2)(11D-27)}{6} +  H^2h^4 \dac{3(D-1)(D-2)^2(D-3)}{4} +
\right. \\ \\ \left. + Hh^5 \dac{(D+1)(D-1)(D-2)(D-3)(D-4)}{12} + \right. \\ \\ \left. +
 h^6\dac{D(D-1)(D-2)(D-3)(D-4)(D-5)}{144} \]
- \Lambda =0
\end{array} \label{h_gen}
\end{equation}

\noindent for a dynamical equation that corresponds to $h$.

Finally, the last metric configuration we are going to consider in this review is the case with the spatial curvature. In this case, the metric takes a {form}

\begin{equation}
\begin{array}{l}
ds^{2}=-dt^{2}+a(t)^{2}d\Sigma_{(3)}^{2}+b(t)^{2}d\Sigma_{({D})}^{2}\ ,  
\end{array} \label{metric3}
\end{equation}%

\noindent where $d\Sigma _{(3)}^{2}$ and $d\Sigma _{({D})}^{2}$ stand for the metric of two constant curvature manifolds $\Sigma _{(3)}$ and $\Sigma_{({D})}$\endnote{We consider {\it ansatz} for spacetime in form of a warped product \mbox{$M_4\times b(t)M_D$}, where $M_4$ is a Friedmann--Robertson--Walker manifold with scale factor $a(t)$ whereas $M_D$ is a $D$-dimensional Euclidean compact and constant curvature manifold with scale factor $b(t)$.}. It is worth pointing out that even a negative constant curvature space can be compactified by making the quotient of the space from a
freely acting discrete subgroup of $O(D,1)$ \cite{wolf}.

The complete derivation of the equations of motion is similar to the previous cases and could be found in~\cite{CGP1, CGP2}; if we use the following rescaling of the coupling constants

\begin{adjustwidth}{-\extralength}{0cm}
\begin{equation}
\alpha =\frac{\left(D+3\right) \left(D+2\right) \left(D+1\right) }{6}c_{0}\ ,\ \ \ \beta =\frac{\left(D+1\right) D\left(D-1\right) }{6}c_{1}\ ,\ \ \ \gamma =\frac{\left(D-1\right) \left(D-2\right)
\left(D-3\right) }{6}c_{2}\ , \label{abc}
\end{equation}%
\end{adjustwidth}

\noindent and the  following notations

\begin{eqnarray}
A_{(1)} &=&\frac{\overset{..}{a}}{a},\ \ \ C=\frac{\overset{.}{a}\overset{.}{%
b}}{ab},\ \ \ B_{(1)}=\frac{\overset{..}{b}}{b}, \notag \\
A_{(2)} &=&\frac{\left[ \gamma _{\left( 3\right) }+\left( \overset{.}{a}%
\right) ^{2}\right] }{a^{2}},\
\ \ B_{(2)}=\frac{\left[ \gamma _{\left( \mathbf{D}\right) }+\left( \overset{.}{b}%
\right) ^{2}\right] }{b^{2}}\label{ABdef}
\end{eqnarray}

\noindent the equations of motion could be written in the following form:

\begin{adjustwidth}{-\extralength}{0cm}
\begin{equation}
\begin{array}{l}
\mathcal{E}_{0} =0\Leftrightarrow 0=\alpha +\beta \left( B_{(2)}+\dac{6}{%
D-1}C+\dac{6}{D\left(D-1\right) }A_{(2)}\right)
+\gamma \left(  B_{(2)}^{2}+\dac{12A_{(2)}B_{(2)}}{\left(
D-2\right) \left(D-3\right) }+\right.  \\ \\
\left. +\dac{24C^{2}}{\left(D-2\right) \left(D
-3\right) }+\dac{12B_{(2)}C}{\left(D-3\right) }+\dac{24A_{(2)}C}{%
\left(D-1\right) \left(D-2\right) \left(D
-3\right) }\right) ,  \label{eq0}
\end{array}
\end{equation}
\begin{equation*}
\mathcal{E}_{i}=0\Leftrightarrow 0=\alpha +\beta \left( B_{(2)}+\frac{
4A_{(1)}}{D\left(D-1\right) }+\frac{2B_{(1)}}{D-1
}+\frac{2A_{(2)}}{D\left(D-1\right) }+\frac{4C}{\left(
D-1\right) }\right) +\gamma \left(  B_{(2)}^{2}+\right.
\end{equation*}%
\begin{equation*}
+\frac{16A_{(1)}C}{\left(D-1\right) \left(D-2\right)
\left(D-3\right) }+\frac{8B_{(2)}C}{D-3}++\frac{
8A_{(1)}B_{(2)}}{\left(D-2\right) \left(D-3\right) }+
\frac{8A_{(2)}B_{(1)}}{\left(D-1\right) \left(D-2\right)
\left(D-3\right) }+
\end{equation*}%
\begin{equation}
\left. +\frac{16B_{(1)}C}{\left(D-2\right) \left(D
-3\right) }+\frac{4B_{(1)}B_{(2)}}{\left(D-3\right) }+\frac{
4A_{(2)}B_{(2)}}{\left(D-2\right) \left(D-3\right) }+
\frac{8C^{2}}{\left(D-2\right) \left(D-3\right) }\right)
\ \ ,  \label{eqq1}
\end{equation}%
\end{adjustwidth}
while the equation $\mathcal{E}_{a}=0$ reads%

\begin{adjustwidth}{-\extralength}{0cm}
\begin{equation*}
\mathcal{E}_{a}=0\Leftrightarrow 0=\frac{D}{\left(D-
4\right) }\alpha +\frac{\left(D-2\right) }{\left(D-
4\right) }\beta \left( B_{(2)}+\frac{6A_{(1)}}{\left(D-1\right)
\left(D-2\right) }+\frac{2B_{(1)}}{D-2}+\frac{6A_{(2)}}{%
\left(D-1\right) \left(D-2\right) }+\right.
\end{equation*}%
\begin{equation*}
\left.+\frac{6C}{\left(
D-2\right) }\right) +\gamma \left( B_{(2)}^{2}+\frac{48A_{(1)}C}{\left(D
-2\right) \left(D-3\right) \left(D-4\right) }+\frac{
12B_{(2)}C}{D-4}+\frac{24C^{2}}{\left(D-3\right) \left(
D-4\right) }+\right.
\end{equation*}%
\begin{equation*}
+\frac{12A_{(1)}B_{(2)}}{\left(D-3\right) \left(D
-4\right) }+\frac{24A_{(2)}B_{(1)}}{\left(D-2\right) \left(
D-3\right) \left(D-4\right) }+\frac{24B_{(1)}C}{\left(
D-3\right) \left(D-4\right) }+\frac{4B_{(1)}B_{(2)}}{%
\left(D-4\right) }+
\end{equation*}%
\begin{equation}
\left. +\frac{12A_{(2)}B_{(2)}}{\left(D-3\right) \left(D
-4\right) }+\frac{24A_{(2)}C}{\left(D-2\right) \left(D-
3\right) \left(D-4\right) }+\frac{24A_{(1)}A_{(2)}}{\left(
D-1\right) \left(D-2\right) \left(D-3\right) \left(
D-4\right) }\right).   \label{eqq2}
\end{equation}
\end{adjustwidth}

This finalizes the list of the models we will use throughout this review. The equations of motion are provided here for the general case (for the arbitrary number of spatial dimensions) and for a vacuum; however, particular cases we will be considering have specified a number of dimensions, so terms of a different nature but the same structure will be summarized and the final equations of motion will look simpler.
 As for the source, we will consider vacuum, cosmological constant ($\Lambda$-term), and perfect fluid; for non-vacuum cases, these equations need to be appropriately modified---density added to the constraint equation and pressure to dynamical equations corresponding to spatial dimensions.

\section{EGB: Vacuum Case}
\label{sec_EGB_vac}

In this section, we review the results for EGB gravity (i.e., maximally with quadratic contribution taken into account) and with no matter source (vacuum model). In this case, the equations of motion are exactly as in (\ref{H_gen}) and (\ref{con_gen}) with $\Lambda = 0$ and 
$\beta = 0$. We shall use $D=1$ as an example of the method and of the features and consider it in more detail. The equations of motion for the $D=1$ vacuum case reads  ($H$-equation, $h$-equation, and constraint, respectively)

\begin{equation}
\begin{array}{l}
4\dot H + 6H^2 + 2\dot h + 2h^2 + 4Hh + 8\alpha \( 2(\dot H + H^2)Hh + (\dot h + h^2)H^2\) = 0,
\end{array} \label{D1_H}
\end{equation}
\begin{equation}
\begin{array}{l}
6\dot H + 12H^2 + 24\alpha (\dot H + H^2)H^2 = 0,
\end{array} \label{D1_h}
\end{equation}
\begin{equation}
\begin{array}{l}
6H^2 + 6Hh + 24\alpha H^3h = 0,
\end{array} \label{D1_con}
\end{equation}

\noindent where $H$ is the Hubble parameter corresponding to the three-dimensional world, $h$ is the Hubble parameter corresponding to extra dimensions, and 
$\alpha$ is Gauss--Bonnet coupling. 

From (\ref{D1_con}), we can easily see that
\begin{equation}
\begin{array}{l}
h = - \dac{H}{1+4\alpha H^2},
\end{array} \label{D1_hh}
\end{equation}

\noindent and so $H$ and $h$ have opposite signs for $\alpha > 0$, but could have the same sign in the $\alpha < 0$ case.  Also, we can resolve (\ref{D1_h}) with respect to $\dot H$ to obtain
\begin{equation}
\begin{array}{l}
\dot H = - \dac{2H^2 (1+2\alpha H^2)}{1+4\alpha H^2};
\end{array} \label{D1_dH}
\end{equation}

\noindent and after that, with the use of (\ref{D1_dH}), one can solve (\ref{D1_H}) to obtain
\begin{equation}
\begin{array}{l}
\dot h = - \dac{2H^2 (8\alpha^2 H^4 +  2\alpha H^2 - 1)}{(1+4\alpha H^2)(16\alpha^2 H^4 + 8\alpha H^2 + 1)}.
\end{array} \label{D1_dh}
\end{equation}

One can see $(1+4\alpha H^2)$ in the denominator of (\ref{D1_hh})--(\ref{D1_dh}); it becomes zero at $H^2 = -1/(4\alpha)$ and it corresponds to what we called nonstandard singularity in the Introduction: $H$ is finite while $h$ and curvature become infinite. Later on, for more complicated cases, we shall see that nonstandard singularities could happen for both $H$ and $h$ being finite, and nonstandard singularities form smooth lines on the $(H, h)$ plane.

\begin{figure}[H]%
\includegraphics[width=0.9\textwidth, angle=0]{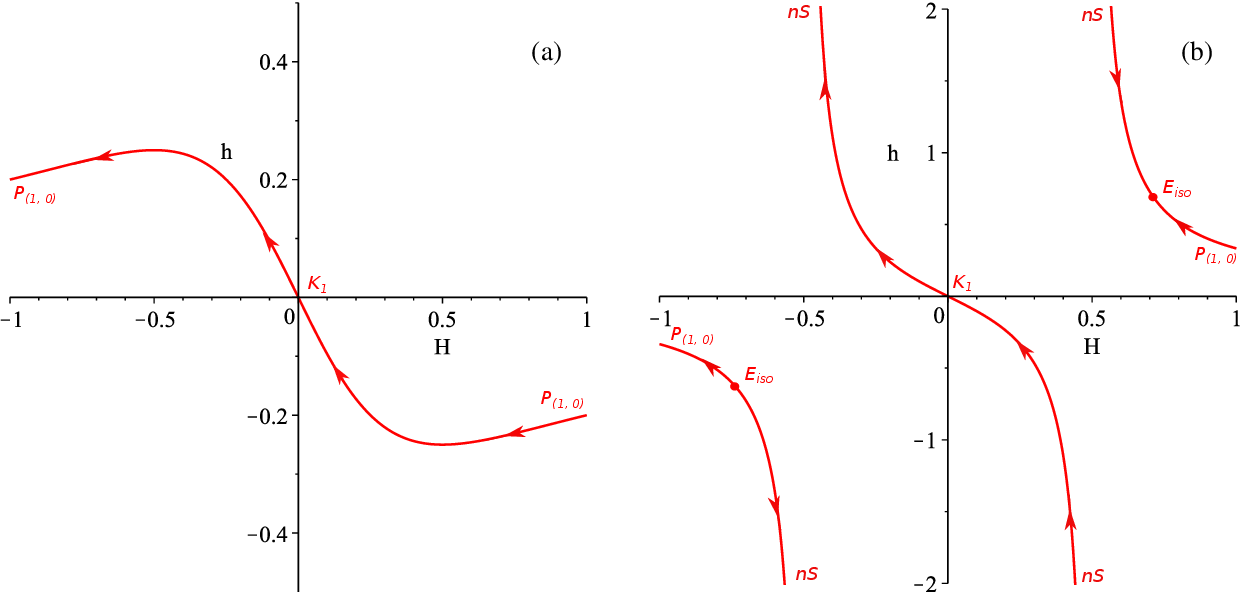}
\caption{Resulting regimes for the $D=1$ vacuum case: $\alpha > 0$ on (\textbf{a}) and $\alpha < 0$ on (\textbf{b})
(see the text for more~details).}\label{D1v}
\end{figure}

For given $\alpha$, Equations (\ref{D1_dH}) and (\ref{D1_dh}) form a system of equations that govern the evolution of the Universe. In~\cite{my16a}, we performed a full analysis of the system with all technical details (see also~\cite{my18a} for another review) while here we just provide the idea: for a given $\alpha$, (\ref{D1_dH}) and (\ref{D1_dh}) could be analyzed to see the asymptotic regimes and stable points and then combined together. The result of this analysis is presented in Figure~\ref{D1v}. On the (a) panel we provided the evolutionary behavior on the $(H, h)$ plane for $\alpha > 0$, while on the (b) panel, for $\alpha < 0$, arrows correspond to the direction of evolution.

Let us have a closer look at Figure~\ref{D1v}a: within the $(H>0, h<0)$ quadrant we have $P_{(1, 0)} \to K_1$ regime transition. These regimes stand for the following: $P_{(1,0)}$ stands for the power-law regime with $p_H = 1$ and $p_h = 0$, where $p_H$ and $p_h$ are Kasner exponents: $p_i = - H_i^2/\dot H_i$ (they characterize the power of power-law behavior $a(t) \propto t^{p}$). Then, $K_1$ is another power-law regime, namely, the Kasner regime in GR~\cite{Kasner} (that is why ``1'' index---in GR for Kasner regime $\sum\limits_p = 1)$. Then, for $\alpha > 0$, we have a regime with $H>0$ (our three-dimensional Universe is expanding) and $h<0$, where extra dimensions are contracting with the Kasner regime as a future asymptote, meaning that the mentioned expansion and contractions are power-law. However, the observed expansion of our Universe is accelerated, which corresponds to the exponential solution. That derived regime could explain the Friedmann stage of the Universe's evolution, but the exponents are different from what is expected in standard cosmology (see~\cite{my16a} for details). Thus, the resulting regime cannot describe our current state of evolution. The other branch, which exists in the $(H<0, h>0)$ quadrant, is just the time-reversal of the just-described transition: $K_1 \to P_{(1, 0)}$.

Now, let us have a closer look at  Figure~\ref{D1v}b and focus on the $H>0$ half-plane (another $H<0$ half-plane is just a time-reversal of the $H>0$ one). This particular example is a plot for $\alpha = - 1$ so that there is a nonstandard singularity (designed as $nS$) at $H^2 = 1/4$ and we clearly see it at $H=1/2$: $h$ diverges both from the left and the right. Then,  for small $H$, we have the $nS \to K_1$ transition, and for larger $H$, there are two: $nS \to E_{iso}$ and $P_{(1, 0)} \to E_{iso}$. The $E_{iso}$ is the isotropic exponential solution (isotropic in an extra-dimensional sense so that all four spatial dimensions are expanding and equal to each other) and obviously it contradicts observations. Then,  we are left with no viable regimes in this case.

Summarizing the $D=1$ vacuum case, for $\alpha > 0$ there is a ``would-be'' viable transition $P_{(1, 0)} \to K_1$, but the late-time asymptote is the Kasner solution, which is not favorable according to observations; for $\alpha < 0$, there are no viable transitions at all.

\begin{figure}[H]%
\includegraphics[width=0.85\textwidth, angle=0]{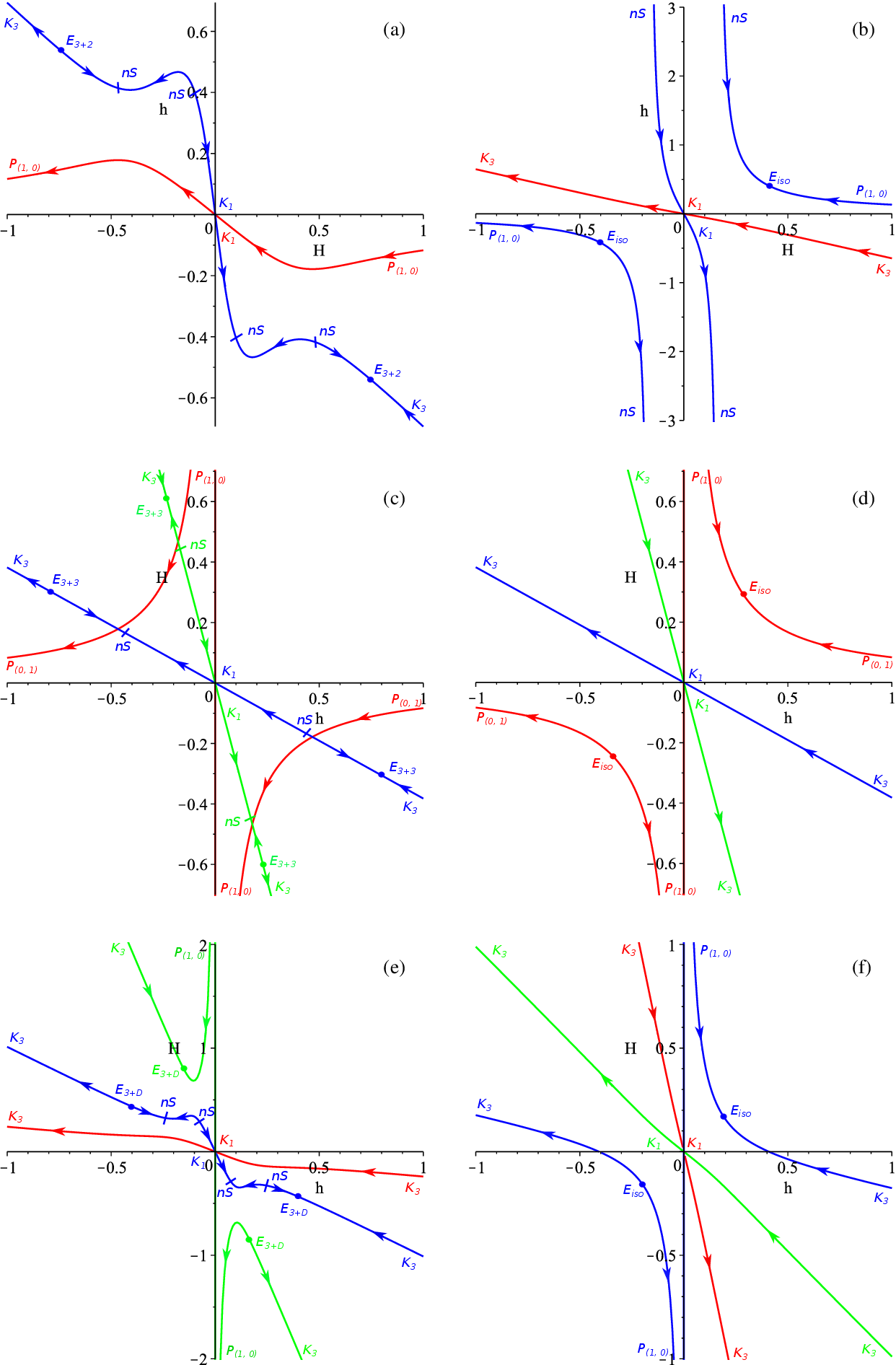}
\caption{{Resulting} 
 regimes for EGB vacuum cases: $D=2$ (panel (\textbf{a}) for $\alpha > 0$  and panel (\textbf{b}) for $\alpha < 0$), $D=3$ (panel (\textbf{c}) for $\alpha > 0$ and panel (\textbf{d}) for $\alpha < 0$), and general $D \geqslant 4$ case (panel (\textbf{e}) for $\alpha > 0$, and panel (\textbf{f}) for $\alpha < 0$)
(see the text for more details).}\label{vac}
\end{figure}

Quite similarly, we perform an analysis for the vacuum model with another $D$: $D=2$, $D=3$, and general $D \geqslant 4$ cases (equations of motion (\ref{H_gen}) and (\ref{con_gen}) have different structures for cases $D=1, 2$, and $3$, while for all $D \geqslant 4$, they remain the same---that is why we consider these cases separately). The difference appears in the number of branches---for $D=1$, the constraint Equation (\ref{D1_con}) has only one root for $h$---Equation (\ref{D1_con}); however, for $D=2$, there will be two, and for $D \geqslant 3$, there will be three; then, the number of branches of solutions will also increase accordingly. The results for the analysis, similar to those presented for $D=1$, are presented in Figure~\ref{vac}. There, on the (a) and (b) panels, we presented the results for $D=2$ ($\alpha > 0$ on (a) and $\alpha < 0$ on (b)), on the (c) and (d) panels---for $D=3$ ($\alpha > 0$ on (c) and $\alpha < 0$ on (d)), and on (e) and (f)---for general $D\geqslant 4$ case ($\alpha > 0$ on (e) and $\alpha < 0$ on (f)); different colors correspond to different branches.

Browsing $\alpha < 0$ cases (panels (b), (d), and (f)), we can conclude that there are no viable regimes---there is a $K_3 \to K_1$ regime (where $K_3$ is GB Kasner with $\sum\limits_p = 3$ (see, e.g.,~\cite{prd09} for the generalization of Kasner regimes on higher Lovelock orders)), but the late-time asymptote is $K_1$. On the contrary, for $\alpha > 0$ and $D\geqslant 2$, there is always a viable transition $K_3 \to E_{3+D}$. There, $E_{3+D}$ stands for an exponential solution with different exponents for three and extra dimensions. Since the solution in question is found in the $(H>0, h<0)$ quadrant, it has expanding three and compacting extra dimensions, which is exactly what we are looking for. Let us note that for $D=3$, since both subspaces are three-dimensional, it does not matter which of them is expanding and which is contracting, as there are two different exponential solutions $E_{3+3}$ that satisfy our conditions. 

Concluding EGB vacuum cases, for $\alpha > 0$ and for all $D\geqslant 2$, there always exists the $K_3 \to E_{3+D}$ transition, which is viable as the late-time regime is exponential with expanding three and contracting extra dimensions. 

Before moving to the next case, let us support our choice of discarding power-law but not exponential regimes: the widely accepted parameter, which quantifies the acceleration of the Universe, is the deceleration parameter defined as

$$
q = - \dac{\ddot a a}{\dot a^2},
$$

\noindent where $a$ is the scale factor. For the exponential {\it ansatz}, the scale factor reads $a(t) = \exp(Ht)$, so that $\dot a = H \exp(Ht)$ and $\ddot a = H^2 \exp(Ht)$, leading to $q_{exp} \equiv -1$, meaning that the exponential solution always expands with acceleration. On the other hand, for the power-law {\it ansatz}, the scale factor reads $a(t) = a_0 t^p$, so $\dot a = p t^{p-1}$ and $\ddot a = p(p-1) t^{p-2}$, resulting in 

$$
q_{power} = - \dac{p(p-1) t^{p-2+p}}{p^2 t^{2(p-1)}} = -1 + \dac{1}{p}.
$$

For accelerated expansion, we need $q < 0$, and for that, $p > 1$ is required. Apparently, all values for $p_H$---power-law exponents corresponding to three-dimensional subspace---originally found in~\cite{my16a}, are $0 < p_H < 1$, so all asymptotic $K_1$ regimes cannot be called realistic. The same is true for $P_{(1, 0)}$, as it has $p_H = 1$, which leads to $q_{power} \equiv 0$, which is still not accelerated expansion. Concluding, exponential solutions are always viable, while power-law solutions could be viable, if $p_H > 1$ which, however, is not the case according to~\cite{my16a}, they are $0 < p_H < 1$. So, hereafter, we consider $E_{3+D}$ realistic compactification regimes while $K_1$ is not viable.

\section{EGB: \boldmath$\Lambda$-Term Case}
\label{sec_EGB_Lambda}

Our next step is to consider EGB gravity but with $\Lambda$-term as a source. The equations of motion are  (\ref{H_gen})--(\ref{con_gen}) but with $\beta = 0$. The equations of motion seem to be slightly different---we only added $\Lambda$ to the right-hand side---but the resulting dynamics change drastically. First of all, there is no more low-energy Kasner regime $K_1$---Kasner regime is a vacuum regime---it approaches $(H=0, h=0)$ in a power-law way, and for vacuum, $(H=0, h=0)$ is a solution (though a trivial solution), as seen from (\ref{con_gen}). For the $\Lambda$-term case, however, r.h.s. of (\ref{con_gen}) is replaced with nonzero $\Lambda$ and so $(H=0, h=0)$ is no longer a solution.

An analysis of all possible transitions also becomes more complicated---instead of only one parameter $\alpha$ in the vacuum case, now we have two---
$\alpha$ and $\Lambda$---and transition availability is now a function of both.

The procedure for the $\Lambda$-term case is the same as for the vacuum case, so we immediately turn to the results; the detailed analysis with all technical details could be found in~\cite{my16b} for $D=1, 2$ and in~\cite{my17a} for $D=3$ and general $D\geqslant 4$ cases; see also~\cite{my18a} for the shorter review version. Analysis suggests that there are no viable regimes for the $D=1$ case, but for $D\geqslant 2$ there are. In particular, for $D=2$, there is a $K_3 \to E_{3+2}$ transition, which exists for $\alpha > 0$ and $\alpha\Lambda < 1/2$ (including the entire $\Lambda < 0$ region). In addition, again for $\alpha > 0$ and $\alpha\Lambda < -1/6$ there is another transition $P_{(1, 0)} \to E_{3+2}$ to the same exponential solution, so for $\alpha > 0$ and $\alpha\Lambda < -1/6$ we have one viable exponential solution $E_{3+2}$, which is a future asymptote, and there could be two different past asymptotes---$K_3$---Gauss--Bonnet Kasner solution and $P_{(1, 0)}$---another power-law solution with $p_H = 1$ and $p_h = 0$; hereafter, we shall denote such a ``double'' transition as $K_3 \to E_{3+2} \ot P_{(1, 0)}$.

For $D=3$, the situation is slightly different---again, similar to the vacuum case, since both subspaces are three-dimensional, it does not matter which of them is expanding and which is contracting, which ``doubles'' the number of viable regimes and, correspondingly, transitions. Thus, for $\alpha < 0$, $\Lambda > 0$, 
$\alpha\Lambda \leqslant -3/2$, there are two ``double transitions'' $K_3^1 \to E_{3+3}^1 \ot P_{(1, 0)}$ and $K_3^2 \to E_{3+3}^2 \ot P_{(0, 1)}$; please mind that $K_3^1$ and $K_3^2$ are two different GB Kasner solutions and $E_{3+3}^1$ and $E_{3+3}^2$ are two different viable exponential solutions. The fact that there are two of them is coming from the fact that both subspaces are three-dimensional. The same two double transitions ($K_3^1 \to E_{3+3}^1 \ot P_{(1, 0)}$ and $K_3^2 \to E_{3+3}^2 \ot P_{(0, 1)}$) also exist for the entire $\alpha > 0$, $\Lambda < 0$, while for $\alpha > 0$, $\Lambda > 0$, $\alpha\Lambda \leqslant 1/2$, they reduce to a couple of transitions $K_3^{1, 2} \to E_{3+3}^{1, 2}$. 

\begin{figure}[H]%
\includegraphics[width=0.85\textwidth, angle=0]{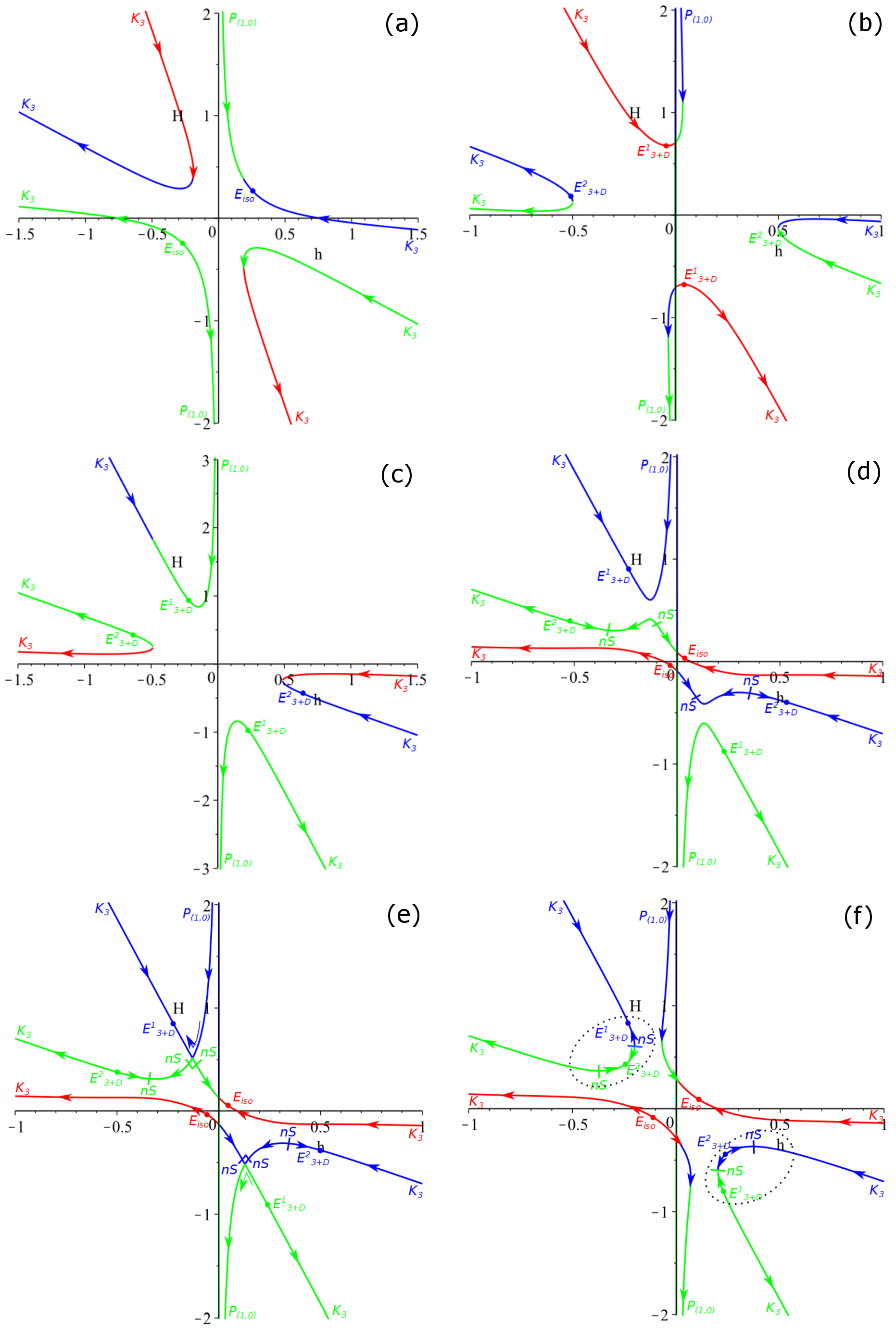}
\caption{Viable regimes for $D \geqslant 4$ EGB cosmology with $\Lambda$-term: (\textbf{a}) panel: $\alpha < 0$, $\Lambda < 0$ featuring $K_3 \to K_3$ regime; (\textbf{b}) panel: $\alpha < 0$, $\Lambda > 0$, $\alpha\Lambda \leqslant \zeta_1$ featuring $K_3 \to E_{3+D}^1 \ot P_{(1, 0)}$ transitions; (\textbf{c})~panel:  $\alpha > 0$, $\Lambda < 0$ featuring the same $K_3 \to E_{3+D}^1 \ot P_{(1, 0)}$ transitions; (\textbf{d}) panel: $\alpha > 0$, $\Lambda > 0$, $\alpha\Lambda < \zeta_2$ still featuring the same $K_3 \to E_{3+D}^1 \ot P_{(1, 0)}$ transitions; (\textbf{e}) panel: $\alpha > 0$, $\Lambda > 0$, $\alpha\Lambda = \zeta_2$ with $K_3 \to E_{3+D}^1 \ot P_{(1, 0)}$ transitions; (\textbf{f}) panel: $\alpha > 0$, $\Lambda > 0$, $\zeta_2 < \alpha\Lambda < \zeta_3$
(see the text for more~details).}\label{GBL}
\end{figure}

Finally, the general $D \geqslant 4$ case has an additional regime---$K_3 \to K_3$---transition between two different GB Kasner regimes. Similar to the $K_3 \to K_1$ transition in the vacuum case, we cannot call it entirely viable, as the future regime is power-law, but the existence of $K_3$ as a future asymptote is interesting; the transition is demonstrated in Figure~\ref{GBL}a.

Similarly to the $D=3$ case, realistic transitions exist in three domains on the $(\alpha, \Lambda)$ space. The first of them is $\alpha < 0$, $\Lambda > 0$ with the additional constraint $\alpha\Lambda \leqslant \zeta_1$; let us note that within this domain, there were no realistic transitions for $D=2$. Realistic transitions here are $K_3 \to E_{3+D}^1 \ot P_{(1, 0)}$---the same realistic exponential solution could be reached from two different high-curvature power-law past asymptotes; these transitions are demonstrated in Figure~\ref{GBL}b.

Mentioned above, $\zeta_1$ was derived in~\cite{my17a} (though denoted as $\zeta_3$ there) and it reads

\begin{equation}
\begin{array}{l}
\zeta_1 = - \dac{D(D-1)}{4(D-2)(D-3)};
\end{array} \label{genD_zeta1}
\end{equation}

\noindent this limit replaces the corresponding $\alpha\Lambda = -3/2$ limit from the $D=3$ case. 

The same transitions exist in another domain as well---in $\alpha > 0$, $\Lambda < 0$, where they exist within the entire domain; this situation is demonstrated in Figure~\ref{GBL}c. 

Finally, the same transitions also exist in the $\alpha > 0$, $\Lambda > 0$ domain, but there, its location and surroundings depend on $\alpha\Lambda$:
for $\alpha\Lambda < \zeta_2$, the situation is depicted in Figure~\ref{GBL}d, while for $\alpha\Lambda = \zeta_2$, it is presented in Figure~\ref{GBL}e. 
For $\alpha\Lambda > \zeta_2$, the situation is presented in Figure~\ref{GBL}f and one can see that the double transition $K_3 \to E_{3+D}^1 \ot P_{(1, 0)}$ degrades to a single $K_3 \to E_{3+D}^1$ for  $\alpha\Lambda > \zeta_2$. Further, for  $\alpha\Lambda > \zeta_3$, $E_{3+D}^1$ is replaced with nonstandard singularity and so no realistic transitions exist for  $\alpha\Lambda > \zeta_3$.

Mentioned above, $\zeta_2$ and $\zeta_3$ are also derived from~\cite{my17a}, where they are denoted as $\zeta_2$ and $\zeta_6$, respectively:

\begin{adjustwidth}{-\extralength}{0cm}
\begin{equation}
\begin{array}{l}
\zeta_2 = \dac{\sqrt[3]{\mathcal{D}_2 (D-1)^2}}{12(D-2)(D-1)D(D+1)} + \\ \\ + \dac{(D^6 - 6D^5 + 10D^4 - 20D^2 + 24D + 36)(D-1)}{3D(D-2)(D+1)\sqrt[3]{\mathcal{D}_2 (D-1)^2}} +
\dac{D^3 - 9D^2 + 8D + 24}{12D(D-2)(D+1)},~\mbox{where} \\ \\
\mathcal{D}_2 = 10D^{10} + 6D^9 \mathcal{D}_1 - 100D^9 - 30D^8 \mathcal{D}_1 + 330 D^8 + 30D^7 \mathcal{D}_1 - 240 D^7 + 54D^6 \mathcal{D}_1 - \\ - 600D^6 - 84D^5 \mathcal{D}_1 + 240D^5 - 24D^4 \mathcal{D}_1 + 1520D^4 +
48D^3 \mathcal{D}_1 + 640 D^3 - 2880D^2 + 1728 \\  \mbox{and}~ \\ \mathcal{D}_1 = \dac{(D-4)(D-3)(D+2)}{(D-1)(D+1)}\sqrt{\dac{(D-4)(D+2)}{D(D-2)}};
\end{array} \label{genD_zeta2}
\end{equation}
\end{adjustwidth}

\begin{equation}
\begin{array}{l}
\zeta_3 = \dac{1}{4} \dac{3D^2 - 7D + 6}{D(D-1)};
\end{array} \label{genD_zeta3}
\end{equation}

\noindent let us note that $\zeta_3$ replaces the corresponding limit $\alpha\Lambda = 1/2$ from $D=2, 3$ cases. 

So, concluding with realistic transitions existing in EGB cosmologies with $\Lambda$-term as a source, starting from $D \geqslant 2$, we always have 
realistic transitions---either double  $K_3 \to E_{3+D} \ot P_{(1, 0)}$ or single $K_3 \to E_{3+D}$; the area of parameters on the $(\alpha, \Lambda)$ space where they exist depends on $D$, but it is always an open region.

\section{Cubic Lovelock: Vacuum Case}
\label{sec_L3_vac}

For this case, we use the same set of Equations (\ref{H_gen}) and (\ref{con_gen}) but with $\Lambda = 0$. The original study is performed in~\cite{cubL1} for $D=3, 4$ and in~\cite{cubL2} for $D=5$ and general $D\geqslant 6$ cases, and here, we summarize our findings.
Similar to the EGB case, depending on $D$, different terms will be nonzero, so we consider all possible $D$ concluding with a general $D$ case where all terms are present. This way, we start with $D=3$, as it is the lowest dimension where cubic Lovelock terms are nontrivial; analysis suggests~\cite{cubL1} that in that case, there are two sets of regimes, $P_{(1, 0)} \to E_{3+3}^1$ and $P_{(0, 1)} \to E_{3+3}^2$, which exist for $\alpha > 0$, $\beta < 0$ and $P_{(1, 0)} \to K_1^1$, $P_{(0, 1)} \to K_1^2$, in turn which exist for $\alpha > 0$, $\beta > 0$.
 We have two sets for the same reason we have the ``doubling'' of regimes in $D=3$ EGB cases because both subspaces are three-dimensional. Let us note that the second regime is Kasner and cannot be called viable for the same reason as for EGB gravity cases. Let us also note that both regimes exist for $\alpha > 0$.
 
The next case to consider is $D=4$ (see~\cite{cubL1} for a full analysis); similar to the $D=3$ case, we have two transitions. The first of them is $P_{(1, 0)} \to E_{3+4}$, which exists for $\alpha > 0$, $\mu \leqslant \mu_1$, where $\mu \equiv \beta / \alpha^2$ and $\mu_1 = -4\sqrt[3]{98}/135 + 38/135 \approx 0.1449$ were found in~\cite{cubL1}. So, the entire $\beta < 0$ is included in the existence region for this transition. Another transition is $P_{(1, 0)} \to K_1$, which exists for $\alpha > 0$, $\mu > \mu_1$. Again, since the low-curvature regime is Kasner, we cannot truly call it realistic. However, it is interesting to note that, similar to the $D=3$ case, both regimes exist only for $\alpha > 0$.

Surprisingly, for $D=5$, we have exactly the same regimes with exactly the same locations but with another $\mu_1$ (see~\cite{cubL2} for details):

$$
\mu_1 = - \dac{\sqrt[3]{2150 + 210\sqrt{105}}}{210} + \dac{2}{21\sqrt[3]{2150 + 210\sqrt{105}}} + \dac{11}{42} \approx 0.1903.
$$

Finally, the case $D \geqslant 6$ is supposed to be general---indeed, for $D \geqslant 6$, all possible terms are present in the equations of motion  (\ref{H_gen}) and (\ref{con_gen}), and so equations of motion for $D \geqslant 6$ contain all possible terms. However, analysis of these equations reveals~\cite{cubL2} that the resulting  regimes are different in $D=6, 7$ and $D \geqslant 8$. 
Realistic regimes in $D=6, 7$ are the same as in $D=3, 4, 5$ but with different values for $\mu_1$: for $D=6$ we have $\mu_1 \approx 0.2151$ while for $D=7$ we have $\mu_1 \approx 0.2324$ (please mind that in the original research~\cite{cubL2} they are denoted differently due to different numbering of special points in different $D$ cases). 

\begin{figure}[H]%
\includegraphics[width=0.85\textwidth, angle=0]{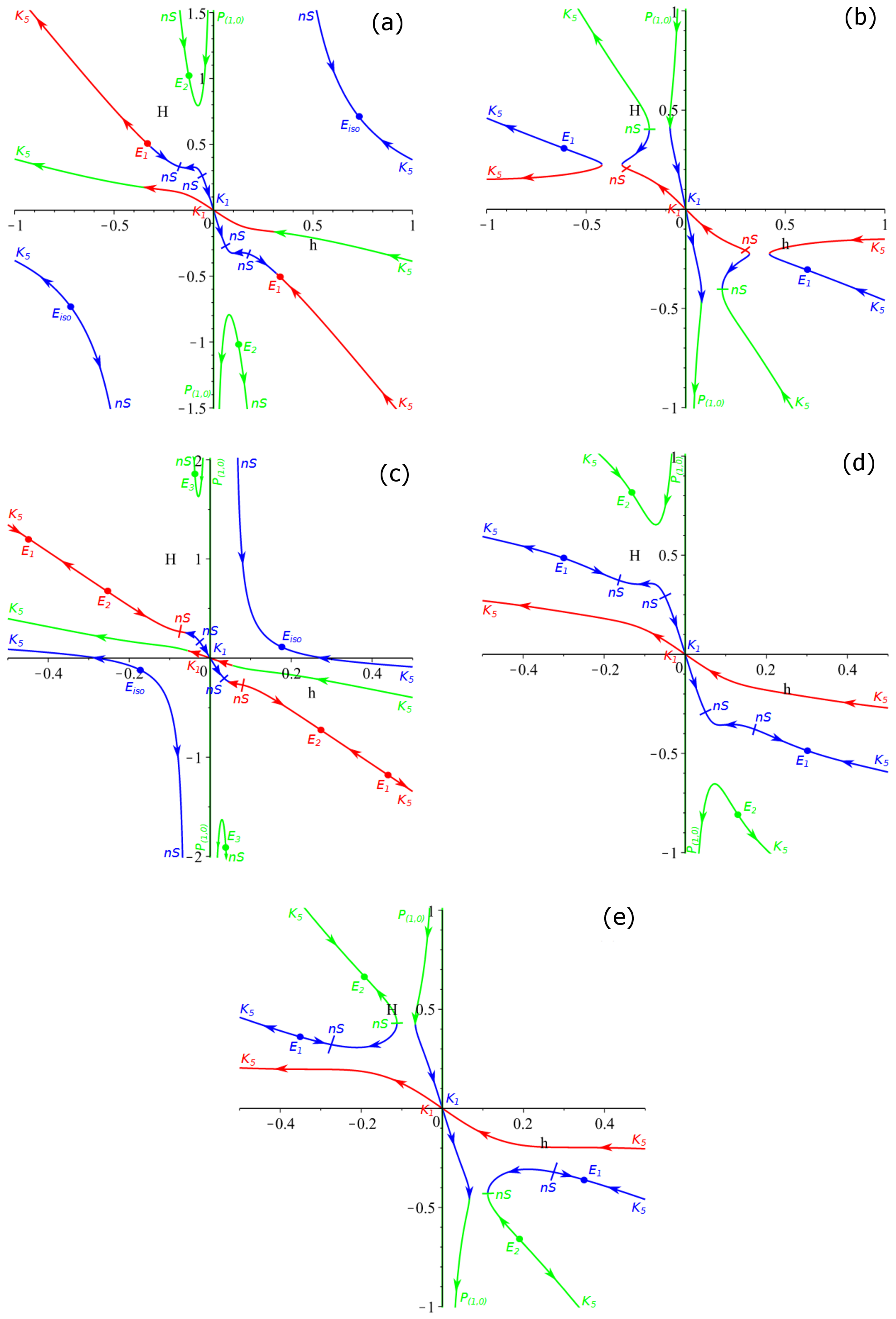}
\caption{Viable regimes for vacuum cubic Lovelock cosmology (all viable regimes are located in the second quadrant): 
(\textbf{a}) panel: $P_{(1, 0)} \to E_{3+D}$ transition on the green branch for $D = 3 \div 7$ and $\alpha > 0$, $\mu \leqslant \mu_1$; 
(\textbf{b}) panel: $P_{(1, 0)} \to K_1$ transition on the green-blue branch for $D = 3 \div 7$ and $\alpha > 0$, $\beta > 0$, $\mu > \mu_1$;
(\textbf{c}) panel: $P_{(1, 0)} \to E_{3+D}$ transition on the green branch as well as $K_5 \to E_{3+D}$ transition on the red branch for $D \geqslant 8$ and $\alpha > 0$, $\beta < 0$, $\mu \leqslant \mu_2$;
(\textbf{d}) panel: $P_{(1, 0)} \to E_{3+D} \ot K_5$ double transition on the green branch for $D \geqslant 8$ and $\alpha > 0$, $\beta > 0$, $\mu \leqslant \mu_3$;
(\textbf{e}) panel: $P_{(1, 0)} \to K_1$ transition on the right green-blue branch and $K_5 \to E_{3+D}$ transition on the left green-blue branch for $D \geqslant 8$ and $\alpha > 0$, $\beta > 0$, $\mu > \mu_3$
(see the text for more details).}\label{cubL}
\end{figure}

The mentioned above regimes for $D = 3 \div 7$ are demonstrated in Figure~\ref{cubL}a---$P_{(1, 0)} \to E_{3+D}$---and in Figure~\ref{cubL}b---$P_{(1, 0)} \to K_1$.

At last, for $D \geqslant 8$, the abundance and structure of the viable solutions changes drastically: for $\alpha > 0$ and $\beta < 0$ there are two realistic compactifications, $P_{(1, 0)} \to E_{3+D}^1$, which exist everywhere in $\beta < 0$, demonstrated in Figure~\ref{cubL}c, and $K_5 \to E_{3+D}^2$, which exists for $\mu \leqslant \mu_2$, where

$$
\mu_2 = - \dac{D^4 + 30D^3 + 189 D^2 - 540 D + 324}{D^4 - 6D^3 - 25D^2 + 102D - 72},
$$

\noindent again found in~\cite{cubL2}; this regime is demonstrated in Figure~\ref{cubL}d. So, for $\mu \leqslant \mu_2$, there exist two different compactification schemes for two different exponential solutions $E_{3+D}^1$ and $E_{3+D}^1$. For $\alpha > 0$, $\beta > 0$, $\mu \leqslant \mu_3$, we have double transition $K_5 \to E_{3+D}^3 \ot P_{(1, 0)}$, demonstrated in Figure~\ref{cubL}d, which splits into $K_5 \to E_{3+D}^3$ and $P_{(1, 0)} \to K_1$ for $\mu > \mu_3$, demonstrated in {Figure}~\ref{cubL}e. The quoted value $\mu_3$ is the lesser root of a certain sixth-order polynomial mentioned in ~\cite{cubL2}.

Concluding with the realistic compactification regimes in vacuum models in cubic Lovelock gravity, we can notice that they exist in all 
$D \geqslant 3$, but their structure and abundance differ for $D = 3 \div 7$ and $D \geqslant 8$: for $D = 3 \div 7$, we need $\mu \leqslant \mu_1$, while for $D \geqslant 8$, realistic compactifications exist for all $\beta$; another interesting feature is that all of them exist only for $\alpha > 0$.

\section{EGB: Perfect Fluid Case}
\label{sec_EGB_perf.fluid}

To study EGB cosmology with perfect fluid as a source, we use Equations (\ref{H_gen}) and (\ref{con_gen}) with $\beta = 0$ and replace $\Lambda$ with $(-p)$ in (\ref{H_gen}) and (\ref{h_gen}) and $\Lambda$ with $\rho$ in (\ref{con_gen}) where $p$ is the pressure and $\rho$ is the density of the perfect fluid. We use $\omega$ for the equation of state which links pressure and density as $p = \omega \rho$. Please mind that $\omega$, being an equation of state, is constant, while $\rho$ is a dynamical variable. Formally, since we have matter in the form of a perfect fluid, the system should be supplemented with a continuity equation, but the full system (\ref{H_gen})--(\ref{con_gen}) plus continuity equation is overdetermined, so we drop the continuity equation; similar techniques are used for the analysis of Friedmann equations with perfect fluid.
Originally, research was performed in~\cite{my18d} and covers only low-dimensional cases ($D=1, 2$), which we are going to review here. 

The $D=1$ case demonstrates the existence of three nonsingular regimes, but neither of them has realistic compactification. In contrast, in
$D=2$, there is a number of nonsingular regimes, but only one of them is realistic---$K_3 \to E_{3+2}$---and it happens for $\alpha > 0$ and $\omega < 1/3$; the measure of the initial conditions leading to this regime increases with the growth of $\omega$, reaching the maximum at $\omega = 1/3 - 0$. 

\begin{figure}[H]%
\includegraphics[width=0.8\textwidth, angle=0]{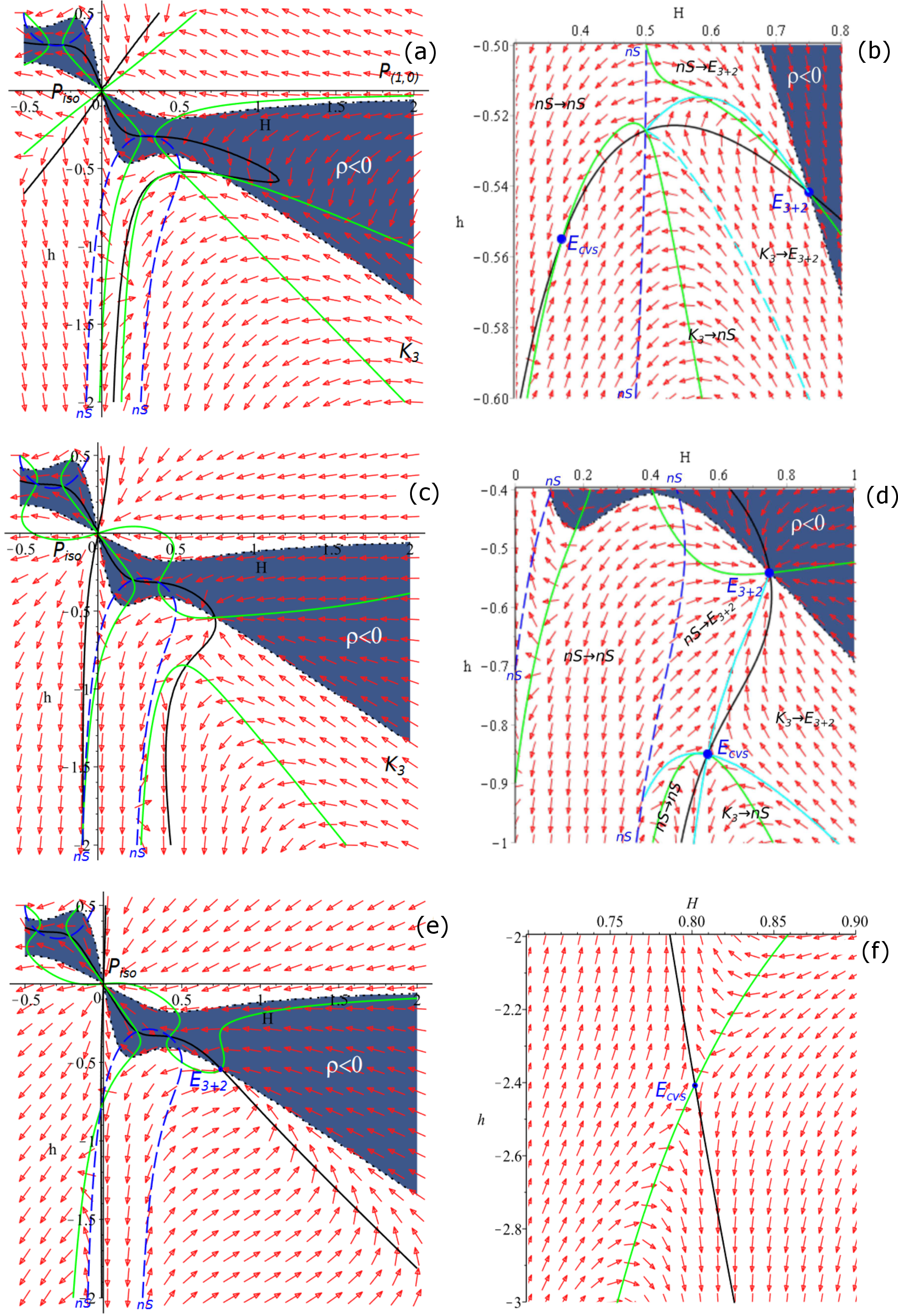}
\caption{Viable regimes for EGB model with a perfect fluid as a source: 
(\textbf{a}) panel: large-scale structure of the $H>0$, $h < 0$ quadrant for $D=2$, $\alpha > 0$, $\omega < 0$; 
(\textbf{b}) panel: vicinity of $E_{3+2}$ stable point for $D=2$, $\alpha > 0$, $\omega < 0$, initial conditions leading to $K_3 \to E_{3+2}$ transition are bounded by light-blue lines;
(\textbf{c}) panel: large-scale structure of the $H>0$, $h < 0$ quadrant for $D=2$, $\alpha > 0$, $1/3 > \omega > 0$;
(\textbf{d}) panel: vicinity of $E_{3+2}$ stable point for $D=2$, $\alpha > 0$, $1/3 > \omega > 0$,  initial conditions leading to $K_3 \to E_{3+2}$ transition are bounded by light-blue lines;
(\textbf{e}) panel: large-scale structure of the $H>0$, $h < 0$ quadrant for $D=2$, $\alpha > 0$, $\omega > 1/3$, only $nS \to E_{3+2}$ transition remains;
(\textbf{f}) vicinity of the exponential constant volume solution ($E_{CVS}$)
(see the text for more details).}\label{perf.fl}
\end{figure}

These regimes are demonstrated in Figure~\ref{perf.fl}: (a) and (b) panels correspond to the $\omega < 0$ case (the (a) panel shows large-scale structure of the $H>0$, $h<0$ quadrant while the (b) panel focuses on the vicinity of $E_{3+2}$ exponential solution), the (c) and (d) panels---to $0 < \omega < 1/3$ (again, the (c) and (d) panels represent large-scale and fine structures, respectively) and the (e) panel---to $\omega > 1/3$ and we can see that $K_3 \to E_{3+2}$ is no more; we have $nS \to E_{3+2}$ instead.

As Figure~\ref{perf.fl} differs significantly from the previous cases, it requires additional explanation. First of all, evolution curves: for vacuum and $\Lambda$-term cases, evolution curves are determined by nonzero parameters of the system; $\alpha$, $\beta$, and $\Lambda$: in this case, we have variable parameter---density $\rho$---so that instead of predetermined evolution curves, we have a field, which is represented by red arrows.
 However, this is a physical parameter, and it has to be strictly positive, but not all $(H, h)$ combinations correspond
to $\rho > 0$. We denoted such ``forbidden'' $\rho < 0$ regions in dark blue. To simplify the analysis of the regimes, we also added $\dot H = 0$ lines in green, and $\dot h = 0$ lines in black so that their intersections have $\dot H = \dot h = 0$, which corresponds to exponential solutions (which are also fixed points of the system). The last piece is the nonstandard singularity; their locations are marked with dashed blue lines.

These notations are enough for analyzing the resulting regimes. Additional numerical simulations reveal particular past or future asymptotes and the vector field demonstrates the general flow from and to these asymptotes. For instance, on the (a) panel, in the bottom part we can see that the past asymptote is $K_3$ (found numerically), and the vector field shows the direction to $nS$, making the transition $K_3 \to nS$; we perform a similar analysis for all other regimes. 

The light-blue lines on  Figure~\ref{perf.fl}b,d are bound initial conditions, which lead to $E_{3+2}$ from $K_3$, so that initial conditions form the only realistic compactification regime $K_3 \to E_{3+2}$. Comparing Figure~\ref{perf.fl}b with Figure~\ref{perf.fl}d, one can notice an increase in this area, so that the measure of the realistic compactification regime is increasing as $\omega$ is growing until it reaches $1/3$; for $\omega > 1/3$, we still have $E_{3+2}$, but $K_3$ is no longer reachable as the field changes direction (compare {Figure~\ref{perf.fl}c} 
 and {Figure~\ref{perf.fl}e}) and it is $nS$ as a past  asymptote, resulting in an $nS \to E_{3+2}$ transition, which is not realistic compactification.

After considering the EGB model with perfect fluid as a source, we can conclude that, similar to previously considered EGB models, realistic compactification exists in \mbox{$D=2$}; we have not considered this model in higher dimensions yet. It is interesting that realistic compactification exists only for $\omega \leqslant 1/3$; we are going to return to this question in the Discussion~section.

\section{EGB: Spatial Curvature Case}
\label{sec_EGB_curv}

\begin{figure}[H]%
\includegraphics[width=1.0\textwidth, angle=0]{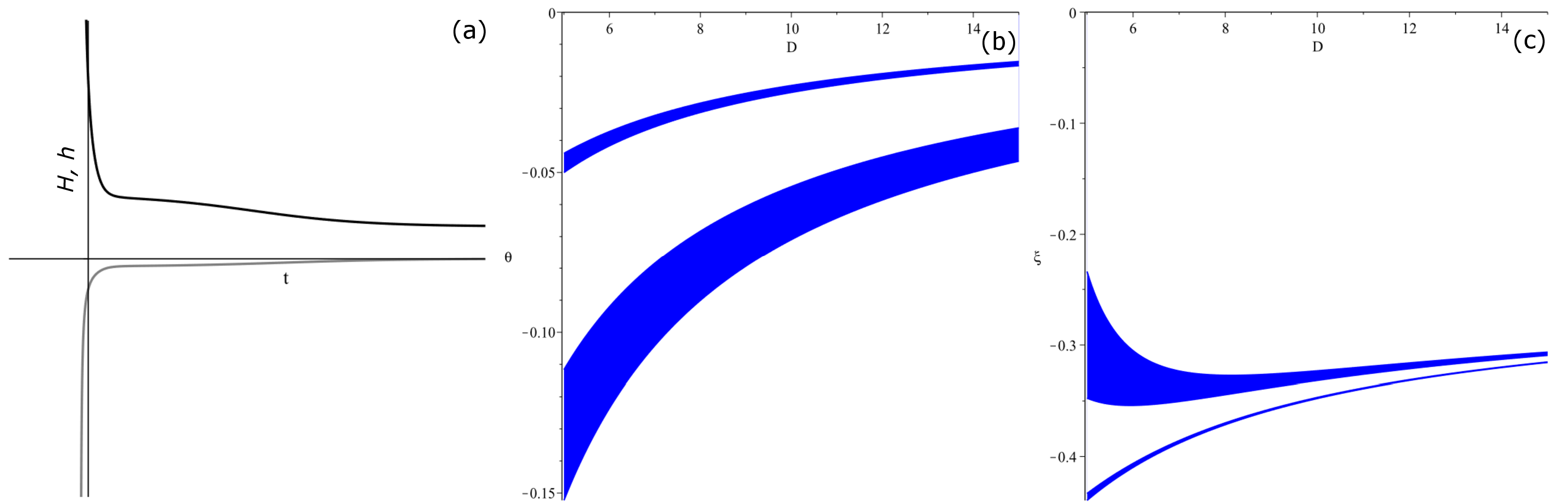}
\caption{Illustrations for the dynamics of the EGB case with spatial curvature: regime with stabilization of extra dimensions ($H > 0$, $h\to 0$) on (\textbf{a}) panel, areas on the parameters space where stabilization of extra dimensions with positive spatial curvature is possible and stable, as a function of $D$ ((\textbf{b},\textbf{c}) panels 
(see the text for more details).}\label{curv}
\end{figure}

Spatial curvature plays an important role in GR: negative curvature typically leads to expansion, while positive could lead to recollapse; particularly, positive curvature could change the measure of the trajectories reaching inflationary asymptotic~\cite{infl1, infl2}. In the context of extra dimensions, the effects of spatial curvature were not investigated much until we found a particular solution where extra dimensions could be stabilized if the curvature of extra dimensions is negative and the parameters are within some non-vanishing range; this situation is illustrated in Figure~\ref{curv}a, where we presented behavior for Hubble parameters and one can see that the 3-dimensional subspace is expanding ($H > 0$) while the size of extra dimensions is stabilized (as $h \to 0$, $b(t) \to \const$). Initially, this regime was described in~\cite{CGP1} by numerically solving (\ref{eq0})--(\ref{eqq2}) for a variety of parameters; with the follow-up investigation performed in~\cite{CGP2}, where we described all other regimes that could appear within the same model, and~\cite{CGPT}, where the influence of matter in the form of perfect fluid on regime availability was studied. In~\cite{PT2017}, we additionally investigated the influence of the spatial curvature of three-dimensional subspace on the dynamics of the regimes. 

Other notable contributions to the study of this regime and its availability and abundance include~\cite{fr1}, where it was demonstrated that in the presence of the cubic Lovelock term, the stabilization of extra dimensions could coexist with isotropization---it is not so in the EGB case. It is also worth mentioning that we observed a similar effect for the vacuum model with a cubic Lovelock contribution---successful compactification schemes could coexist even with two different isotropizations---a situation we do not have in vacuum EGB (see Sections~\ref{sec_EGB_vac} and~\ref{sec_L3_vac}).
 In addition to this contribution, we can also mention~\cite{fr3}, where a certain case is investigated numerically, and~\cite{fr4}, where the $(3+7)$-dimensional case in the presence of the cubic Lovelock contribution is investigated numerically. 

Finally, the stability of the extra dimensions stabilization regime is investigated in~\cite{our20}. There, we build a system of the perturbation equations based on (\ref{eq0})--(\ref{eqq2}) and perturbed around the exact solution $H \equiv \dot a(t)/a(t) = \const = H_0$, $b(t) = \const = b_0$. The results of the investigation suggest that for negative curvature of extra dimensions, the resulting regime with stabilization of extra dimensions always exists and is stable. However, for many reasons, the case of positive curvature is more attractive---it is more obvious how to make extra dimensions compact in the case of positive spatial curvature---so we kept looking for the same regime but with positive curvature. 

The results of our investigations are as follows: for $D=2$, there formally exist solutions with the stabilization of extra dimensions, but they are unstable. On the contrary, starting from $D=3$, there are stable solutions of the desired type for both positive and negative curvatures of extra dimensions. This way, for $D=3$, for positive curvature, stable solutions exist for $\alpha < 0$ and $\xi \in (-0.5448, -0.5) \cup (0, +\infty)$ (where $\xi = \alpha\Lambda$), and thus for entire $\Lambda < 0$ and the small positive range of $\Lambda$. 
 For negative curvature of extra dimensions in $D=3$, stable solutions exist for $\alpha < 0$, $\xi < -3/2$; it is interesting to note that both of them exist only for $\alpha < 0$. Next, for $D=4$, stable solutions for positive curvature of extra dimensions exist for $\alpha < 0$, $\xi \in (-27/54, -15/32) \cup (-0.3, 3/8)$---again, for both signs of $\Lambda$, but now both ranges are finite. For negative curvature, there are two domains with different conditions: one is the same as for $D=3$: $\alpha < 0$, $\xi < -3/2$ and the other is $\alpha > 0$, $\Lambda < 0$. 
Finally, for the general $D \geqslant 5$ case, the situation is as follows: for positive curvature of extra dimensions, stable solutions exist for $\alpha < 0$, $\xi \in (\xi_1, \xi_2) \cup (\xi_3, \xi_4)$ (see Figure~\ref{curv}b,c where we provide this range in $\theta = \alpha H_0^2$ and $\xi$ variables). For negative curvature, one of the domains is the same as before, $\alpha < 0$, $\xi < -3/2$, while another is $\alpha > 0$, $\xi < - D(D-1)/(4(D-2)(D-3))$. 

It is interesting to note that for positive curvature, we require $\alpha < 0$, while for negative $\alpha$, both could be signs for $D\geqslant 4$.

Overall, the stabilization of extra dimensions could take place in all $D \geqslant 3$: for the negative curvature of extra dimensions, it is a natural process, while for positive curvature, it requires fine-tuning the parameters, and this fine-tuning becomes more and more tight with the growth of $D$.

\section{EGB: General Anisotropic Case}
\label{sec_EGB_aniso}

\begin{figure}[H]%
\includegraphics[width=1.0\textwidth, angle=0]{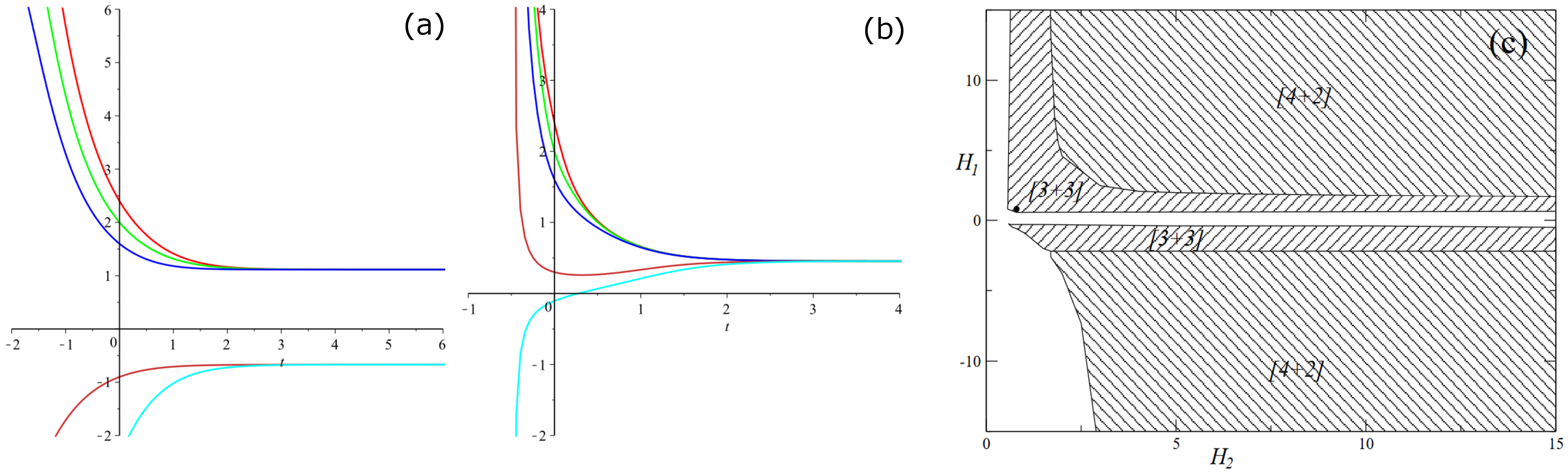}
\caption{Illustrations for the dynamics of the EGB case initial total anisotropy (Bianchi-I-type): different initial conditions for the model with 5 spatial dimensions could lead to either $[3+2]$ spatial splitting ((\textbf{a}) panel) or isotropization ((\textbf{b}) panel); distribution of the initial conditions for the model with 6 spatial dimensions leading to either $[4+2]$ or $[3+3]$ spatial splittings ((\textbf{c}) panel)
(see the text for more details).}\label{aniso}
\end{figure}

Last but not least, the case to report here is the case where all spatial dimensions are initially totally anisotropic, making it a Bianchi-I-type configuration. In this case, the equations of motion are (\ref{constr_gen})--(\ref{dyn_gen}) with $n=2$ maximum. Then,  we start with some initial conditions and integrate the equations to see the past and future asymptotic regimes. For the case with four spatial dimensions, the investigation was performed in~\cite{prd10} for vacuum, $\Lambda$-term, and perfect fluid cases. However, since there are no realistic compactifications for $D=1$ neither in vacuum nor in $\Lambda$-term or perfect fluid cases, the only nonsingular regime reported there is isotropization. By singular regimes, we mean the situation when evolution is interrupted by nonstandard singularity, defined the same as before, but for totally anisotropic cases, nonstandard singularities are not points on the evolutional curve but are surphases in the phase space and are quite abundant (see, e.g.,~\cite{prd10} for the simplest case with four spatial dimensions). 
The analysis was further continued in~\cite{PT2017} for five spatial dimensions, where we know compactification for $D=2$ exists, presented in Figure~\ref{aniso}a, while isotropization for the same case is presented in Figure~\ref{aniso}b. Since there are no exponential solutions with other spatial splittings in five spatial dimensions, these two are the only possible future asymptotes. However, in six spatial dimensions, the situation is different, as there could be two different principal spatial splittings: $[4+2]$, with expanding four and contracting two dimensions, and $[3+3]$ with expanding three and contracting another three dimensions. Of these two splittings, only the latter is realistic, but we can end up in both starting from the initially anisotropic case, which is demonstrated in Figure~\ref{aniso}c. There, we plot a two-dimensional slice of the six-dimensional Hubble parameter space with the distribution of the initial conditions, leading to either $[3+3]$ splitting or $[4+2]$. One can see that the area for $[4+2]$ is larger; however, it is a two-dimensional slice and to judge the real measure, one needs to take into account the distribution across all six dimensions.

In addition to the mentioned references in~\cite{fr2}, the additional numerical analysis of the five and six spatial dimensional cases was performed. As in~\cite{PT2017}, we performed an analysis of both the anisotropic model and the curvature, and we proposed a scheme that would allow us to address the question of reaching the stabilization of the extra dimensions regime from an initially anisotropic state. It cannot be performed directly (as the initial state is Bianchi-I and it is spatially flat while the stabilization is achieved if there is a spatial curvature) so we developed an indirect scheme: we first address the possibility of reaching $[3+D]$ spatial splitting from the initial anisotropy for some parameters, then address the possibility of reaching compactification from the initial conditions with initially very small but nonzero curvature. So, in~\cite{fr4}, this scheme was tested for seven spatial dimensions: initially, the possibility of forming $[3+4]$ spatial splitting was addressed, and after that, the possibility of building the stabilization of extra dimensions from $[3+4]$ spatial splitting with initially small negative curvature.

Summarizing, an initially anisotropic Universe in the course of evolution would end up in one of the spatial splittings, available for a given number of spatial dimensions, or encounter nonstandard singularity. For the former case, we always have an isotropic solution as a possibility (though it exists not for all values of the parameters) and, starting with five spatial dimensions, we have nontrivial spatial splitting into two (or more) isotropic subspaces, again, subject to existence and stability. Among them, we are interested in configurations with an expanding three-dimensional subspace and contracting other subspaces (all other subspaces, if more than one; however, due to high computational difficulty we have not investigated cases where more than one extra-dimensional subspace exists).

 So, our investigation reveals that in five spatial dimensions, everything is good, while in higher dimensions, other spatial splitting exists and it could be that the Universe ends up with ``wrong'' spatial splitting. It will not be three-dimensional subspace which would expand, resulting in the conclusion that the higher the number of spatial dimensions, the less likely that we will end up with ``correct'' compactification.

\section{Summary and Discussion}

Now, let us summarize and discuss the results reported in this review. The vacuum EGB model is the simplest to consider and it has realistic regimes. For $\alpha < 0$ (negative Gauss--Bonnet coupling), the $K_3 \to K_1$ transition exists, from high-curvature (Gauss--Bonnet) Kasner regime to low-curvature standard (GR) Kasner. Despite having the desired behavior of the scale factors, (expansion of the three- and contraction of extra-dimensional subspaces) the resulting regime is power-law and the resulting exponents do not fit the Friedmann stage, so that, despite being potentially viable, it still does not fit the observations. For $\alpha > 0$, the counterpart of this regime is $P_{(1, 0)} \to K_1$---the same future but different past asymptotes. For this case, it is another power-law regime with $p_H = 1$ and $p_h = 0$. Since it also has $\sum p_i = 3$, it could be mistaken for GB Kasner, but it is a different regime. In a way, it is closer to the Taub solution~\cite{Taub}, but the exact relationship between these two is not established.

Apart from these regimes, there is another: for $\alpha > 0$ and for all $D\geqslant 2$, the $K_3 \to E_{3+D}$ transition always exists, which is realistic---three dimensions expand exponentially while extra dimensions exponentially contract; exponential expansion of three-dimensional subspace agrees with the observed accelerated expansion of the Universe.

The next case to consider is the EGB model with the $\Lambda$-term. This model is more complicated, both in dynamics and regime/transition abundance, compared to the vacuum model. We again have a power-law transition there; this time it is the $K_3 \to K_3$ transition between two different high-curvature (GB) Kasner regimes. Again, they are power-law and cannot be~realistic. 

Realistic transitions for the model with $\Lambda$-term include either double $K_3 \to E_{3+D} \ot P_{(1, 0)}$ or single $K_3 \to E_{3+D}$---in both cases, the late-time asymptote is an exponential solution with expanding three and contracting extra dimensions. These transitions are always present for 
$D \geqslant 2$, but exact ranges of the parameters depend on $D$: 

\begin{itemize}

\item For $D=2$, they exist only for $\alpha > 0$: for $\alpha\Lambda < 1/2$ (including the entire $\Lambda < 0$) the $K_3 \to E_{3+2}$ transition exists, and for $\alpha\Lambda < -1/6$, additional $P_{(1, 0)} \to E_{3+2}$ transitions to the same exponential solution exist, making it a double transition $K_3 \to E_{3+2} \ot P_{(1, 0)}$;

\item For $D=3$, we additionally have transitions for $\alpha < 0$, $\Lambda > 0$,  $\alpha\Lambda \leqslant -3/2$. There, we have two double transitions $K_3^1 \to E_{3+3}^1 \ot P_{(1, 0)}$ and $K_3^2 \to E_{3+3}^2 \ot P_{(0, 1)}$; the same double transitions exist for the entire $\alpha > 0$, $\Lambda < 0$, while for $\alpha > 0$, $\Lambda > 0$, $\alpha\Lambda \leqslant 1/2$, they reduce to a couple of single transitions $K_3^{1, 2} \to E_{3+3}^{1, 2}$;

\item The general $D \geqslant 4$ case has the same regimes as $D=3$, but the limits of their existence are different and $D$-dependent. Double transitions in the negative $\alpha$ domain exist for $\alpha < 0$, $\Lambda > 0$, $\alpha\Lambda \leqslant \zeta_1$, and for positive $\alpha$ single $K_3 \to E_{3+D}$, transitions exist for $\alpha\Lambda < \zeta_3$, while for $\alpha\Lambda \leqslant \zeta_2$, they are supplemented with $P_{(1, 0)} \to E_{3+D}$ transitions and form double transitions. 

\end{itemize}

So, this analysis demonstrates how the usage of simple requirements for the existence of realistic Universe evolution can help us impose constraints on the parameters of the theory. However, Gauss--Bonnet gravity is used not only in cosmology. There are various aspects of Gauss--Bonnet gravity in AdS spaces, such as shear viscosity to entropy ratio, causality violations, or CFTs in dual gravity description, which allow us to put constraints on the parameters as well; these
constraints could be summarized as follows~\cite{alpha1, alpha2, alpha3, alpha4, alpha5, alpha6, alpha7, alpha8}:

\begin{equation}
\begin{array}{l}
 - \dac{(D+2)(D+3)(D^2 + 5D + 12)}{8(D^2 + 3D + 6)^2} \equiv \eta_2 \leqslant \alpha\Lambda \leqslant \eta_1 \equiv \dac{(D+2)(D+3)(3D + 11)}{8D(D+5)^2}.
\end{array} \label{alpha_limit}
\end{equation}

Limits for dS ($\Lambda > 0$) are less numerous and are based on different aspects (causality violations, perturbation propagation, and so on) of black hole physics in dS spaces. The most stringent constraint coming from these considerations is~\cite{BHGB7, BHL2, add_rec_23}

\begin{equation}
\begin{array}{l}
\alpha\Lambda \geqslant \eta_3 \equiv - \dac{D^2 + 7D + 4}{8(D-1)(D+2)}.
\end{array} \label{alpha_limit2}
\end{equation}

\begin{figure}[H]%
\includegraphics[width=1.0\textwidth, angle=0]{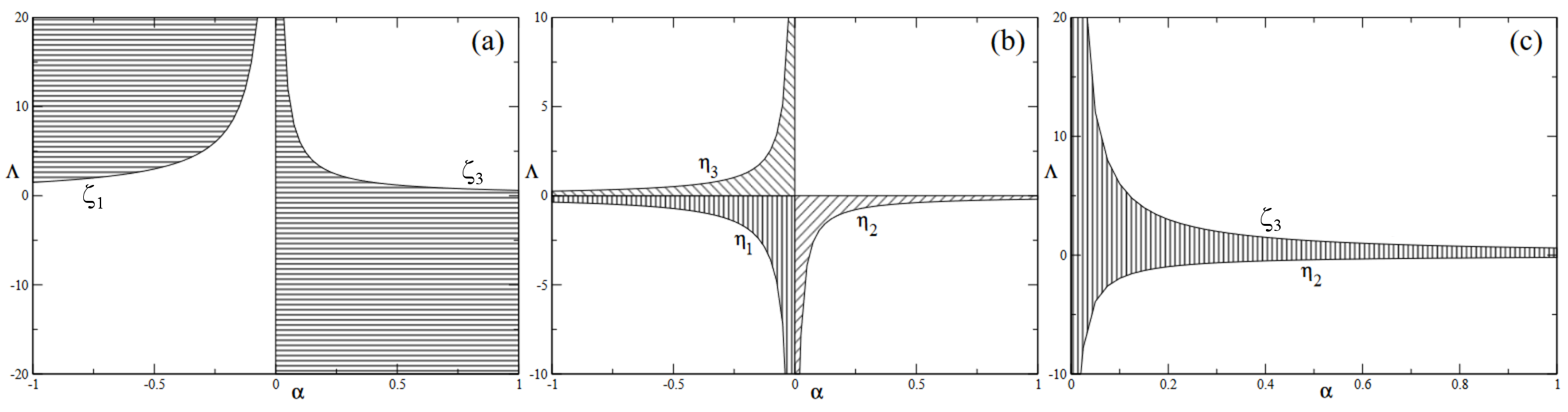}
\caption{Summary of the bounds on $(\alpha, \Lambda)$ from this paper alone on (\textbf{a}) panel; from other considerations found in the literature on (\textbf{b}) panel; and the intersection between them on (\textbf{c}) panel
(see the text for more details).}\label{comp}
\end{figure}

Then,  we can confront our limits with those obtained from other considerations of GB: we put our constraints in Figure~\ref{comp}a (with $\zeta_{1, 2, 3}$ being defined in (\ref{genD_zeta1})--(\ref{genD_zeta3})) and mentioned constraints from the other literature in Figure~\ref{comp}b ($\alpha > 0$, $\Lambda > 0$, which is not shaded but is included) and their intersection is presented in Figure~\ref{comp}c.
 From there, one can see that $\alpha > 0$ is favored by the combination of constraints and the resulting constraint on $\alpha\Lambda$ reads
\begin{equation}
\begin{array}{l}
\dac{3D^2 - 7D + 6}{4D(D-1)}  \equiv \zeta_3 \geqslant \alpha\Lambda \geqslant \eta_2 \equiv - \dac{(D+2)(D+3)(D^2 + 5D + 12)}{8(D^2 + 3D + 6)^2}.
\end{array} \label{alpha_limit3}
\end{equation}

It is worth mentioning that initially this result was obtained in~\cite{my17a}, but as it is an important milestone, we decided to reproduce it here.

The next model we considered is the vacuum cubic Lovelock gravity. Analysis suggests that for all $D \geqslant 3$ cases, there are realistic compactification schemes, though the regime abundance and structure is different in $D = 3 \div 7$ and in $D \geqslant 8$. For $D = 3 \div 7$, there is one realistic transition $P_{(1, 0)} \to E_{3+D}$, which exists for $\alpha > 0$, $\mu \leqslant \mu_1$ (including entire $\beta < 0$); the other ``would-be'' viable transition is $P_{(1, 0)} \to K_1$, but the corresponding late-time asymptote $K_1$ is Kasner regime. It is power-law and its expansion rate is insufficient to describe the current expansion of the Universe;
these regimes are demonstrated in Figure~\ref{cubL}a,b.
For $D \geqslant 8$, we have the following realistic compactifications: for $\alpha > 0$, $\beta < 0$, we have $P_{(1, 0)} \to E_{3+D}$, which exists everywhere in $\beta < 0$ and $K_5 \to E_{3+D}$, which exists for $\mu \leqslant \mu_2$; for $\alpha > 0$, $\beta > 0$, there is double transition $K_5 \to E_{3+D} \ot P_{(1, 0)}$, for $\mu \leqslant \mu_3$ which ``decouples'' into $K_5 \to E_{3+D}$ and $P_{(1, 0)} \to K_1$ (the latter not being quite viable) for $\mu > \mu_3$; these regimes are demonstrated in Figure~\ref{cubL}c--e. 
This way, realistic compactification exists only for $\alpha > 0$, but for all $\beta$, there is at least one viable compactification regime.

Another model we consider is the EGB model, with perfect fluid as a source. Due to complexity, so far we have considered only low-dimensional cases ($D=1,2$), and these are the cases we are reporting here. So, in $D=1$, there are three nonsingular transitions, but none of them have realistic late-time regimes. On the contrary, for $D=2$, there is realistic compactification $K_3 \to E_{3+2}$, and it is achieved for $\alpha > 0$, $\omega < 1/3$ with the measure of the trajectories experiencing this transition, increasing as $\omega \to 1/3 - 0$; the situation is demonstrated in Figure~\ref{perf.fl}.

The model with perfect fluid has two interesting features we want to highlight for our readers. The first of them is what we call ``constant volume solution''---exponential solution with constant volume, defined as follows in the continuity equation:

\begin{equation}
\begin{array}{l}
\dot \rho + \rho (1 + \omega) \sum\limits_i H_i = 0,
\end{array} \label{continuity}
\end{equation}

\noindent where we used $p = \omega \rho$ equation of state; for exponential solutions $H_i \equiv \const$, so that l.h.s. of the equations of motion are constants, and so the density $\rho \equiv \const$, then the continuity Equation (\ref{continuity}) reduces to 

\begin{equation}
(\rho+p)\sum_i H_i=0 \iff
\left[\begin{array}{r}
        \rho=0\quad\mathbf{(a)} \vspace{.2cm}\\
        p=-\rho\quad\mathbf{(b)} \vspace{.2cm}\\
        \sum_k H_k=0\quad\mathbf{(c)}
      \end{array}\right.
\label{continuity2}
\end{equation}

Then,  we can clearly see all three possibilities when exponential solutions could exist---it is either vacuum (case (a)), $\Lambda$-term (case (b)), or the third possibility with $\sum_k H_k=0$ (case (c)). One can see that in the third possibility, the sum of Hubble parameters is zero, so the comoving volume is constant, that is why we call these exponential solutions ``constant volume solutions''. We have a separate paper dedicated to the study of their properties~\cite{CST2} (see also~\cite{CPT1} for the general scheme and~\cite{CPT3} for non-constant-volume solutions) and here we notice their appearance on the actual evolution curves. Well, ``appearance on the evolution curves'' is not quite correct---they appear, but not on the evolution curves---they are ``avoided'' by evolution flux. In Figure~\ref{perf.fl}f, we presented the vicinity of the constant-volume solution and one can see that it is a saddle point. Since it
has $\sum_k H_k=0$, it is not formally stable in either $t\to +\infty$ or $t\to -\infty$ (that would require either $\sum_k H_k>0$ or $\sum_k H_k<0$; see~\cite{iv16}), so it is natural that it is a saddle point. The important point is that, despite the fact that we described these solutions in~\cite{CST2}, their actual appearance and nature remained unknown, so the fact that we discovered their presence in a model with perfect fluid as a saddle point is an important result. 

The second feature we want to highlight for the readers is the fact that for $\rho \to 0$, we obtain vacuum EGB behavior, reported in Section~\ref{sec_EGB_vac}. Indeed, from Figure~\ref{perf.fl}b,d, one can notice that $E_{3+2}$ is located on the $\rho = 0$ boundary, and it is quite clear why. From (\ref{continuity2}), we can see that there are only three possibilities for building an exponential solution---option (c) gives us CVS, and (b) is not an option since we do not have the $\Lambda$-term here, so the only option remaining is (a), and $E_{3+2}$ is located exactly on $\rho = 0$. Moreover, taking the $\rho = 0$ slice from the evolution vector field would reconstruct vacuum EGB behavior, so that the vacuum EGB model is the $\rho\to 0$ limit of the perfect fluid model. While this is somewhat expected, it is still an interesting fact.

The less expected and equally interesting fact is the boundary on the equation of state---the $K_3 \to E_{3+2}$ transition takes place only for $\omega < 1/3$, while for $\omega > 1/3$, it is replaced by $nS \to E_{3+2}$. The explanation lies in the difference in scalings for matter and curvature and $\omega = 1/3$ is a critical value for the GB term (see~\cite{grg10} for details). For contraction (it defines the stability of the past asymptote, and $K_3$ here is a past asymptote), if $\omega < 1/3$, curvature dominates, and GB Kasner is preserved, for $\omega > 1/3$, matter dominates and GB Kasner is no more. That is what is happening with $K_3$ as a past asymptote. 

The next case to consider is the case with the spatial curvature in EGB gravity. After the discovery of the regime, where the scale factor of the extra dimensions could be stabilized~\cite{CGP1}, in EGB gravity, if the spatial curvature of the extra-dimensional section is negative, we started to investigate this regime and the models with nonzero spatial curvature in more detail. Indeed, the regime is interesting, but negative curvature is hard to handle, while positive curvature is more associated with compact manifolds. So, if there was a similar regime but with positive curvature, this would be ideal. Indeed, we found this regime in the case of positive curvature, but its abundance is much less than that of the negative curvature. What is more, its abundance is decreasing with an increase in the number of extra dimensions. Thus, it exists and is stable, but to end up in this regime, we require some sort of fine-tuning.

And finally, the totally anisotropic case---the case where initially we had all dimensions equally distributed (Bianchi-I-type). All previous cases assumed that all spatial dimensions are divided into subspaces---isotropic three-dimensional which represent our Universe, and isotropic extra-dimensional. However, in the most natural scenario, we should start with totally anisotropic space, and it should naturally evolve into a configuration with two isotropic subspaces---this evolution is what we hope to find. The results suggest that for the cases where stable exponential solutions exist, such transitions are possible. For instance, for four spatial dimensions, the only stable exponential solution is isotropic, so only it could be reached. However, in five spatial dimensions, there are two, isotropic and $[3+2]$ splitting---with expanding three and contracting two dimensions---so that the final configurations could be reached (see Figure~\ref{aniso}a,b).
 Similarly, in higher numbers of dimensions, more different spatial splittings are allowed, which makes it less probable to end up with the ``correct'' spatial splitting.

Finally, let us put together all cases where realistic compactification to the exponential solution is happening, and so, $\Lambda$CDM could be recovered in a three-dimensional subspace. All such cases are summarized in Table~\ref{tab1}. There, we indicate the type of model and the source of matter in the first column, the number of extra dimensions in the second column, and the conditions for the parameters under which compactification occurs.

\begin{table}[H]
\caption{All cases with realistic compactification/$\Lambda$CDM recovery.\label{tab1}}
	\begin{adjustwidth}{-\extralength}{0cm}
		\newcolumntype{C}{>{\centering\arraybackslash}X}
		\begin{tabularx}{\fulllength}{CCc}
			\toprule
			\textbf{Model}	& \textbf{Number of Extra Dimensions}	& \textbf{Parameters}    \\
			\midrule
EGB, vac & $D \geqslant 2$ & $\alpha > 0$ \\
 \midrule
\multirow[m]{6}{*}{EGB, $\Lambda$} & $D=2$ & $\alpha\Lambda < 1/2$ (incl. $\Lambda < 0$) \\ \cmidrule{2-3}
                                                          & \multirow[m]{2}{*}{$D=3$}  & $\alpha < 0$, $\alpha\Lambda \leqslant -3/2$ \\
                                                             & &                                             $\alpha > 0$, $\alpha\Lambda < 1/2$ (incl. $\Lambda < 0$) \\ \cmidrule{2-3}
                                                          & \multirow[m]{2}{*}{$D\geqslant 4$} & $\alpha < 0$, $\alpha\Lambda \leqslant \zeta_1$ \\                                               
                                                            & &                                                        $\alpha > 0$, $\alpha\Lambda < \zeta_3$ (incl. $\Lambda < 0$) \\ 
 \midrule
\multirow[m]{2}{*}{cubic Lovelock, vac} & $D = 3 \div 7$ & $\alpha > 0$, $\mu < \mu_1$ \\
                                                             &   $D \geqslant 8$ & $\alpha > 0$ \\
\midrule
EGB, perfect fluid & $D=2$ & $\alpha > 0$, $\omega < 1/3$ \\
\midrule
\multirow[m]{7}{*}{EGB, curvature} & $D \geqslant 3$  &  $\gamma_D < 0$, $\alpha < 0$, $\alpha\Lambda < -3/2$  \\ \cmidrule{2-3}
                                                       & $D=3$ & $\gamma_D > 0$, $\alpha < 0$, $\alpha\Lambda \in (-0.5448, -0.5) \cup (0, +\infty)$ \\ \cmidrule{2-3}
                                                       & \multirow[m]{2.5}{*}{$D=4$} & $\gamma_D > 0$, $\alpha < 0$, $\alpha\Lambda \in (-27/54, -15/32) \cup (-0.3, 3/8)$ \\ \cmidrule{3-3}
                                                                                                  & & $\gamma_D > 0$, $\alpha > 0$, $\Lambda > 0$ \\ \cmidrule{2-3}
                                                        & \multirow[m]{2.5}{*}{$D\geqslant 5$} & $\gamma_D > 0$,  $\alpha < 0$,  $\alpha\Lambda \in (\xi_1, \xi_2) \cup (\xi_3, \xi_4)$ (see Figure~\ref{curv}c) \\ \cmidrule{3-3}
                                                                                                                 & & $\gamma_D < 0$, $\alpha > 0$, $\alpha\Lambda < - D(D-1)/(4(D-2)(D-3))$ \\                                                                                                                      
			\bottomrule
		\end{tabularx}
	\end{adjustwidth}	
\end{table}

\section{Conclusions}

To conclude, we have demonstrated that it is possible to set constraints on the parameters of Lovelock gravity using generic requirements, such as the existence of a nonsingular transition from a high-curvature to a realistic low-curvature regime. Since the theory is multidimensional, ``realistic'' here stands for expanding three- and contracting or static extra-dimensional subspaces. These results are obtained under the condition that the total space is split into two isotropic subspaces---three- and extra-dimensional with the former representing our Universe while the latter is the ``hidden'' extra dimensions. Our results also suggest that the transition from a totally anisotropic (Bianchi-I-type) Universe to the configuration with two isotropic subspaces is quite natural and happening for quite a wide range of parameters. However, in a high number of spatial dimensions, compactification could happen to a different spatial splitting, e.g., six spatial dimensions could end up both as $(3+3)$---three contracting and three expanding---which favors observations, or as $(4+2)$---four expanding and two contracting---which obviously contradicts; the exact outcome depends on the parameters and initial conditions. 

This review reflects the current status of the Lovelock cosmology exploration. From it, one can see that despite some progress, we still are quite far from fully understanding the dynamics of Lovelock cosmology. Most of the progress is conducted within EGB gravity, which is the lowest non-GR contribution of Lovelock gravity. The results reported for cubic Lovelock gravity demonstrate that we cannot extrapolate results for EGB on higher-order Lovelock gravity, we need to investigate at least cubic Lovelock gravity to see if higher orders could be extrapolated from EGB and cubic contributions. So far we investigated only vacuum regimes in cubic Lovelock gravity, and some progress was made on exponential solutions and solutions with spatial curvature, but a full investigation of the regimes in cubic Lovelock gravity with $\Lambda$-term is still ongoing. It is more complicated than the cases considered in this review and it has more parameters, so both the structure of the regimes' abundance and their distribution over parameter space is much more complicated---that slows the investigation, but we are still hoping to complete it in the future.

However, even within EGB gravity, we lack a full understanding and full description of all regimes and cases---vacuum and $\Lambda$-term transitions are described while the model with the perfect fluid as a source still lacks full description. We described low-$D$ cases, which demonstrate a promise in explaining successful compactification in the presence of ordinary matter but are still struggling with the general high-$D$ case. Still, we are planning to complete it in the future, which will broaden our understanding of the dynamics of EGB cosmologies. 

Overall, as just stated, the results reported here demonstrate promise in describing our observed Universe as a three-dimensional subspace of the higher-dimensional theory utilizing Lovelock gravity, which exists only in higher dimensions. As the results demonstrated promise, we are going to continue our investigation of the model and the pursuit of new exciting features and regimes that this beautiful theory could provide.

\vspace{6pt} 
\funding{This research received no external funding.}

\conflictsofinterest{The authors declare no conflicts of interest.}

\begin{adjustwidth}{-\extralength}{0cm}				
\printendnotes[custom]			
\end{adjustwidth}

\begin{adjustwidth}{-\extralength}{0cm}

\reftitle{References}

\PublishersNote{}
\end{adjustwidth}

\end{document}